\documentclass[aps,reprint,nofootinbib,floatfix,pre]{revtex4-2}
\usepackage[a4paper, total={17cm, 25.4cm}]{geometry}
\usepackage{amssymb}
\usepackage{amsmath}
\usepackage{mathtools}
\usepackage{xfrac}
\usepackage{esint}
\usepackage{graphicx} 
\usepackage{tensor} 
\usepackage{physics} 
\usepackage{hyperref} 
\usepackage{caption}
\usepackage{cleveref} 
\usepackage{tikz}
\usepackage{xcolor}
\usepackage{diagbox}
\usepackage{slashbox}
\usepackage{float}

\makeatletter
\@for\next:={int,iint,iiint,iiiint,dotsint,oint,oiint,sqint,sqiint,
  ointctrclockwise,ointclockwise,varointclockwise,varointctrclockwise,
  fint,varoiint,landupint,landdownint}\do{%
    \expandafter\edef\csname\next\endcsname{%
      \noexpand\DOTSI
      \expandafter\noexpand\csname\next op\endcsname
      \noexpand\ilimits@
    }%
  }
\makeatother

\newcommand{\wcircle}[2][black,fill=white]{\tikz[baseline=-0.5ex]\draw[#1,radius=#2] (0,0) circle ;}%
\newcommand\nb{\addtocounter{equation}{1}\tag{\theequation}}

\newcommand\fsing{f_{\textrm{sing}}}
\newcommand\freg{f_{\textrm{reg}}}
\newcommand\fsub{f_{\textrm{sub}}}
\newcommand\fctm{f_{\textrm{ctm}}}
\newcommand\ghigh{G_{\textrm{high}}}
\newcommand\glow{G_{\textrm{low}}}

\newcommand\Phiimh{\Phi_{\textrm{imh}}}
\newcommand\gly{G_{\text{YL}}}
\newcommand\myl{M_{\text{YL}}}
\newcommand\cyl{C_{\text{YL}}}
\newcommand\fyl{f_{\text{YL}}}

\newcommand\beq{\begin{equation}}
\newcommand\eeq{\end{equation}}
\newcommand\bea{\begin{align}}
\newcommand\eea{\end{align}}
\newcommand{\s}{\sigma}
\newcommand{\be}{\beta}

\catcode`,\active

\catcode`\,12

\begin{document}

\title{Corner Transfer Matrix Approach to the Yang-Lee Singularity in the 2D Ising Model in
a magnetic field}
\author{Vladimir V. Mangazeev}
\email{Vladimir.Mangazeev@anu.edu.au}
\author{Bryte Hagan}
\email{Bryte.Hagan@anu.edu.au}
\author{ Vladimir V. Bazhanov}
\email{Vladimir.Bazhanov@anu.edu.au}

\affiliation{Department of Fundamental and Theoretical Physics,
The Australian National University, Canberra, ACT 2601, Australia}
\begin{abstract}
We study the two-dimensional (2D) Ising model in a complex magnetic field in the vicinity of
the Yang-Lee edge singularity. By using
Baxter's variational corner transfer matrix method combined with analytic techniques,
we numerically calculate the scaling function
and obtain an accurate estimate of the location of the Yang-Lee singularity.
The existing series expansions for susceptibility of the 2D Ising model
on a triangular lattice by Chan, Guttmann, Nickel, and Perk 
allowed us to substantially enhance the
accuracy of our calculations. Our results are in excellent agreement with
the Ising field theory calculations by Fonseca, Zamolodchikov 
and the recent work by Xu and Zamolodchikov. 
In particular, we numerically confirm an agreement between
the leading singular behavior of the scaling function and the predictions
of the ${\mathcal M}_{2/5}$ conformal field theory.\\ \\
DOI:\href{https://doi.org/10.1103/PhysRevE.108.064136}{\color{blue}10.1103/PhysRevE.108.064136}

\end{abstract}

\maketitle

\section{Introduction}
The two-dimensional (2D) Ising model plays a
prominent role in the development of the theory of phase transition
and critical phenomena
\cite{O44,Y52,McCoyWu,Bax07,AF83,BPZ84,Zam89a,CGNP11,FZ03,XuZ22}.
In Refs. \cite{MBBD09,MBBD10} the scaling and universality of
the 2D Ising model in a magnetic field were studied by Baxter's corner transfer matrix approach
\cite{Bax07}.
%
%
Here we further continue this study. We
consider the planar nearest-neighbour Ising model on
regular square and triangular lattices.
Its partition function reads
\beq
{Z}=\sum_{\sigma}\exp \Big\{\,\beta
\sum_{\langle ij\rangle} \sigma_i \sigma_j +{H}\sum_i
\sigma_i\,\Big\}\ , \quad \sigma_i=\pm1,\label{Z-def}
\eeq
where the
first sum in the exponent is taken over all edges, the second sum --- over all
sites and the outer sum --- over all spin configurations $\{\s\}$ of the
lattice. The constants $H$ and $\beta=J/k_BT$ denote the (suitably
normalized) magnetic field and inverse temperature.  The free
energy, magnetization and magnetic susceptibility are defined as
\beq
f=-\lim_{N\to\infty}\frac{1}{N}\log Z,\quad M=-\frac{\partial
  f}{\partial H},\quad \chi=-\frac{\partial^2 f}{\partial
  H^2}\ ,\label{F-def}
\eeq
where $N$ is the number of lattice
sites and derivatives are taken at zero field $H=0$.
The model exhibits a second order phase transition at $\beta=\beta_c$ and
$H=0$, where
\beq \beta_c^{(s)}={\textstyle{\frac{1}{2}}}
\log(1+\sqrt{2}),\qquad \beta_c^{(t)}= {\textstyle{\frac{1}{4}}}\log
3,\label{betac}
\eeq
for the square \cite{O44} and triangular \cite{Newell1950}
lattices, respectively.
Since we investigate the large lattice limit, boundary conditions
become irrelevant
in the off-critical regime. We shall be using the fixed boundary
conditions, which
are naturally arising in the corner transfer matrix method.

It is convenient to introduce a new temperature-like variable $\tau$
\beq
{\tau = \frac12\left(\sqrt{k}- \frac1{\sqrt k}\right)},\quad k=(\tau+\sqrt{1+\tau^2})^2 \label{tauk}
\eeq
defined via the ``elliptic modulus'' parameter $k$. The latter is connected
to the inverse temperature $\beta$ through the following
lattice-dependent formulas \cite{CGNP11}
\begin{align}
    k^{(s)} &=\frac{1}{s^2},\quad  s = \sinh 2\beta^{(s)}, \label{lab0} \\
    k^{(t)} &= \frac{4t^{3/2}}{(1-t)^{3/2}(1+3t)^{1/2}},\quad  t = e^{-4\beta^{(t)}},\label{lab1}
\end{align}
where the superscripts $(s)$ and $(t)$ stand for the square and triangular lattices, respectively.

The temperature variable $\tau$ is inspired by the Kramers-Wannier  duality transformation
$\tau\to-\tau$ and $k\to1/k$ around the critical point
$\tau=0$. The high temperature regime corresponds to $\tau>0$ and $k>1$.

Close to the critical point $\tau,H\to 0$,
the singular part of the lattice free energy, $\fsing$,
dominates the free energy $f$ due to a logarithmic singularity.
It has the form
\begin{align}
    \fsing(\tau, H) &= \frac{m^2}{8\pi}\log[m^2] + m^2G(\xi), \>\> \xi=\frac{h}{|m|^{15/8}}, \label{lab5}
\end{align}
where $m$ and $h$ are specially constructed nonlinear scaling
variables \cite{AF80} whose leading terms for small $\tau,H\to 0$ are
proportional to $\tau$ and $H$, respectively, see Eq. \eqref{lab4} below.
The universality implies
that the scaling function $G$ depends on $h$
and $m$ only via the ratio $\xi$ defined in Eq. \eqref{lab5}.  Analytic
properties of the scaling function $G(\xi)$ in the complex plane $\xi$
have been thoroughly analized in Ref. \cite{FZ03}.

In this paper, we investigate a detailed structure of $G(\xi)$ in the
neighborhood of the so-called Yang-Lee edge singularity
\cite{YL52,LY52}.
In particular,
we give a very accurate estimate for the location of
this singularity, extending
the recent results of Ref. \cite{XuZ22}.


The paper is organised as follows. In Sec. \ref{secscal} we give a brief
exposure of the Ising scaling theory, discuss different critical
regimes and describe our approach. We also explain the benefits of working
with a triangular lattice instead of the square one.  In Sec. \ref{sectria} we
review known analytic results for the Ising model at zero magnetic field on
a triangular lattice.
We use them to
determine certain parts of the free
energy in the thermodynamic limit. The calculations of the
magnetic susceptibility are reviewed in Sec. \ref{secsus}.
In Sec. \ref{secCTM}
we explain the corner
transfer matrix method and its modification for the triangular
lattice. In Sec. \ref{secnum}
we present all numerical results which we  used to
determine the expansion of the scaling function near the Yang-Lee
singularity. Finally, in Conclusion we discuss the results and make some
comments on unsolved problems.

\section{Scaling theory\label{secscal}}

According to the scaling theory, the lattice free energy \eqref{F-def}
in the vicinity of the critical point $\tau,H\to 0$ can be written as
\begin{align}
    f(\tau, H) &= \fsing(\tau, H)+ \freg(\tau, H) + \fsub(\tau, H),
    \label{lab2}
\end{align}
where the leading singular part
\beq
\fsing(\tau, H)=\mathcal{F}(m(\tau,H),h(\tau,H)), \label{lab3}
\eeq
can be expressed through a universal
scaling function $\mathcal{F}(m,h)$,
with the parameters $\tau$ and $H$ entering the right-hand side (RHS) of Eq. \eqref{lab3}
only through nonlinear scaling
variables \cite{AF80} 
\begin{align}
    m(\tau, H)&=-C_\tau\,a(\tau)\,\tau +
\mu_h \,H^2b(\tau)+ \mathcal O(H^4),\nonumber \\
   h(\tau, H)&=C_H\,c(\tau)\,H + e_h\,d(\tau)\,H^3 + \mathcal O(H^5). \label{lab4}
\end{align}
Note, that all the constants and coefficient
functions in the above formulas are lattice dependent. Here we assume
the following normalization
\begin{align}
&a(\tau)=1+O(\tau), \quad b(\tau)=1+O(\tau),\nonumber\\  &c(\tau)=1+O(\tau),
\quad d(\tau)=1+O(\tau). \label{lab40}
\end{align}
with $C_\tau>0,\,C_H>0$.
The leading singular part \eqref{lab3} 
has the form 
\begin{align}
\mathcal{F}(m,h)    &= \frac{m^2}{8\pi}\log[m^2] + m^2G(\xi), \>\> \xi=\frac{h}{|m|^{15/8}}. \label{lab5a}
\end{align}
which was already quoted in Eq. \eqref{lab5}.
Next, the regular part $\freg(\tau,H)$ in Eq. \eqref{lab2} contains no
singularities at $\tau,H\to 0$ and have the following Taylor expansion
\begin{align*}
    \freg(\tau, H) &= A(\tau) + H^2B(\tau) + \mathcal O(H^4).\label{lab6}
\nb
\end{align*}
The last term $\fsub(\tau, H)$ in Eq. \eqref{lab2} denotes the
less singular than $\fsing(\tau, H)$ part. It can be written
as
\beq
\fsub(\tau, H)=H^2\fsub(\tau) +O(H^4).\label{lab7}
\eeq
The first nontrivial contribution
to $\fsub(\tau)$, associated with irrelevant operators in the Ising
CFT \cite{Cas02},
will be determined from the susceptibility
calculations of ref. \cite{CGNP11}.

Analytic properties of the scaling function $G(\xi)$ for complex $\xi$
have been discussed in details in Refs. \cite{FZ03,XuZ22}.
In fact, to describe $\mathcal{F}(m,h)$ for both positive and negative
$m$ it is convenient to use two different (but, of course,
related) scaling functions in Eq. \eqref{lab5a}.
In
the high-temperature regime $m<0$ the scaling function
$G(\xi)$ admits a series expansion
\beq
\ghigh(\xi)=G_2\xi^2+G_4\xi^4+G_6\xi^6+G_8\xi^8+\ldots.\label{lab8}
\eeq
convergent in some domain around the origin $\xi=0$. Note, that
this expansion contains only even powers of $\xi$, since for $m<0$
there is a symmetry $H\to -H$.

In the low-temperature regime $m>0$, the scaling function admits an
asymptotic expansion for small $\xi$
\beq
\glow(\xi)=\tilde G_1\xi+\tilde G_2\xi^2+\tilde G_3\xi^3+\tilde G_4\xi^4+\ldots.\label{lab9}
\eeq
containing all powers of $\xi$.
Coefficients $G_i$ and $\tilde G_i$ are universal for all lattices
\cite{MBBD10}. The first few of these coefficients
are known with a very high accuracy \cite{MBBD09}.
For later use,
in Table \ref{tablef} we quote the coefficients $G_{n}$ with $n\le 10$.
\begin{table}[h]
\caption{Values of coefficients $G_{n}$}
\centering
\begin{tabular}{ll}
\hline\hline
$G_2$ \phantom{aa}\rule{0.0cm}{.4cm}& $ -1.84522807823(1)$ \\
$G_4$ & $\phantom{+}8.3337117508(1)$ \\
$G_6$ & $-95.16897(3)$ \\
$G_8$ & $\phantom{+}1457.62(3)$ \\
$G_{10}$\phantom{|} & $-25891(2)$ \\
 \hline\hline
\end{tabular}
\label{tablef}
\end{table}

Next, it is convenient to introduce another scaling variable
\beq
\eta=\frac{m}{h^{8/15}} \label{lab10}
\eeq
and rewrite Eq. \eqref{lab5a} as
\beq
\mathcal{F}(m,h)=\frac{m^2}{8\pi}\log[m^2] + h^{16/15}\Phi(\eta).\label{lab11}
\eeq
Then for the high-temperature regime one obtains
\beq
\Phi(\eta)=\eta^2\ghigh((-\eta)^{-15/8})
 \quad \mbox{for real}\quad \eta<0.\label{lab12}
\eeq
The function $\Phi(\eta)$ admits the expansion for small $\eta$ \cite{FZ03}
\beq
\Phi(\eta)=-\frac{\eta^2}{8\pi}\log\eta^2+\sum_{k=0}^\infty \Phi_k\eta^k,\label{lab13}
\eeq
where the series converges in a finite domain around
the origin of the complex $\eta$-plane.
The first two coefficients $\Phi_0$ and $\Phi_1$ are known exactly,
thanks to the integrability of
the critical Ising field theory in a magnetic field \cite{Zam89a}.
The higher coefficients $\Phi_k$, with $k\le8$ are known numerically,
see Refs. \cite{FZ03,MBBD09}.

From the high-temperature regime the function $\ghigh(\xi)$
can be analytically continued
to complex values of $\xi$. The Yang-Lee theory \cite{YL52,LY52}
guarantees analyticity of $\ghigh(\xi)$
in the complex $\xi$-plane with two
branch cuts $(-i\infty,-i\xi_0)$ and $(i\xi_0,+i\infty)$  on the
imaginary axis.
These cuts arise from a condensation
of the Yang-Lee zeros of the partition function in the thermodynamic
limit \cite{MUS2017}.
In this paper, we study the neighborhood of the Yang-Lee edge
singularity at $\xi=\pm i\xi_0$, which is another critical point of the Ising
model \eqref{Z-def}
associated with the nonunitary minimal CFT model ${\cal M}_{2/5}$
with the central charge $c=-22/5$ \cite{Cardy85,CM1989}, often
named as ``Yang-Lee'' CFT (YLCFT).
It has only one
primary field $\phi$ with the scaling dimension $2\Delta=-2/5$.  To
analyse this critical point, one needs to consider the ${\cal M}_{2/5}$
CFT perturbed by the relevant
operator $\phi$ together with an infinite number of irrelevant operators. Further
details are given in Sec. \ref{secnum}.

The suggested numerical value of $\xi_0$ in Ref. \cite{FZ03} is
\beq
\xi_0=0.18930(5).\label{lab14}
\eeq
In this paper, we numerically calculate the scaling function
to study its expansion near the
singularity and, in particular,
significantly improve the above estimate for $\xi_0$, see Eq. \eqref{num8}
below.

Our strategy is as follows. We consider the lattice
Ising model in the high-temperature
regime with a complex magnetic field $H$ in the neighborhood of the
Yang-Lee edge singularity,
 $\xi=\pm i\xi_0$, with $\xi_0$ given by Eq. \eqref{lab14}. Using the
numerical corner transfer matrix algorithm, we calculate the lattice free
energy $\fctm(\tau,H)$ with relatively high accuracy (at least
15 decimal places) for
a large set of numerical values of the temperature and magnetic field.

Using Eqs. \eqref{lab2} and \eqref{lab5a}, one can express
the scaling function as
\begin{align}
&G(\xi)=\frac{1}{m^2}\big(\fctm(\tau,H)-\frac{m^2}{8\pi}\log[m^2]\nonumber\\
 &- \freg(\tau, H) - \fsub(\tau, H)\big),\label{lab15}
\end{align}
where all functions in the RHS are given by series in $\tau$ and $H$
and can be determined from analytic
results and numerical calculations in the high/low-temperature regimes.

Consider the high-temperature regime $m<0$ and fix the variable $\xi$.
Then using Eq. \eqref{lab5} the scaling variable $h$ can be written as
\beq h(\tau,H)=\big(-m(\tau,H)\big)^{15/8}\,\xi . \label{lab16}
\eeq
Combining this with Eq. \eqref{lab4} one can invert the series in $H$ and
obtain the following expansion 
\beq
H(\tau)=\xi\,\tau^{15/8}\,\sum_{k=0}^\infty \xi^{2k}\, \tau^{11 k/4}
\,
p_k(\tau),\label{lab17}
\eeq
where $p_k(\tau)$ are regular in $\tau$. Then it is easy to see that the
coefficients of a series expansion of
$H(\tau)$ in powers of $\tau$ are finite polynomials in $\xi$. Thus,
the variable $\xi$ can be assigned to any fixed value ---
not necessarily small.

Substituting Eq. \eqref{lab17} into the RHS of Eq. \eqref{lab15} one obtains
a function of two variables $\xi$ and $\tau$, which
should not depend on $\tau$. This is
a highly nontrivial self-consistency test that determines the accuracy
of the calculation of $G(\xi)$. Indeed, taking two different values
of $\tau$ with a fixed $\xi$, one
should get the same value of $G(\xi)$.

The CTM method can be applied to regular square, triangular, or
honeycomb lattices. As was shown in Refs. \cite{Cas02,CGNP11}, for the Ising
model the CTF
irrelevant operators contribute in the order $\tau^4$ for the square
lattice and in the order $\tau^6$ for the triangular/honeycomb
lattices. This implies that one can achieve much better accuracy for
the triangular lattice \cite{MBBD10}. For this reason, we restrict all
further analysis to this case.

The precision of CTM calculations is limited by the machine
accuracy, which in our case is $10^{-16}$.
As we shall see below, for the triangular lattice one can
reliably control the series expansion coefficients in
the RHS of Eq. \eqref{lab15} up to the order of $\tau^8$. Therefore, there
is an upper limit on the value of $\tau$ for which the unknown higher
terms in Eq. \eqref{lab15} do not affect the precision of calculations.
Having this in mind, one needs, nonetheless,  to
choose $\tau$ as large as possible to speed up the convergence
of the CTM algorithm. The latter is mainly determined by the variable
\beq
\Delta\beta=\beta_c^{(t)}-\beta=\textstyle\frac{1}{4}\tau+O(\tau^2).\label{lab18}
\eeq
Numerical experiments
suggest that the optimal value of $\Delta\beta$
for the maximum precision calculations should be in the range
\beq
0.006\le \Delta\beta \le 0.012\,,
\eeq
where the lower bound is near the boundary
of convergence of the CTM algorithm.
To reach the numerical accuracy of $10^{-15}$
the calculations were performed with corner transfer matrices of the
size $N=170$.

Since the RHS of Eq. \eqref{lab15} contains an extra factor $1/m^2$, the
accuracy of calculations of $G(\xi)$ is reduced by four or five decimal places.
Therefore, one can expect the accuracy of the scaling
function to be around $10^{-10}$. Indeed, we calculated $G(\xi)$  for
a large set of points $\xi_i$ in the complex plane with three values of
$\Delta\beta$,
\beq
\Delta\beta=0.006,\ 0.007,\ 0.012\,,
\eeq
corresponding to
\beq
\tau\simeq 0.024,\ 0.028,\ 0.048\,.\label{tauval}
\eeq
A difference between the values of $G(\xi)$ for these values of $\tau$
never exceeded $2.\times 10^{-10}$. Therefore, it suggests that this
is the accuracy of our calculations of the scaling function $G(\xi)$.

\section{Ising model with $H=0$ on a triangular lattice\label{sectria}}

In this and the next sections we will use all available exact and
perturbation theory results for the 2D lattice Ising model
to determine the lattice-dependent regular \eqref{lab6} and subleading
\eqref{lab7}
contributions to the free energy, as well as to find coefficients
entering the nonlinear scaling variables \eqref{lab4} to
highest possible orders in the variables $\tau$ and $H$.

The free energy of the Ising model on the
triangular lattice can be written in the form of the following integral
\cite{Newell1950}
\begin{align}
&f^{(t)}(\tau,0)=-\frac{1}{2}\log(4\sinh2\beta)-
\frac{1}{8\pi^2}\iint\limits_0^{2\pi} d\phi_1 d\phi_2\nonumber\\
&\times\log[r(\be)-\cos\phi_1-\cos\phi_2+\cos(\phi_1+\phi_2)],\label{part1}
\end{align}
where
\beq
r(\be)=\frac{3+e^{8\be}}{2(e^{4\be}-1)} \label{part2}
\eeq
and
$\be$ and $\tau$ are related through Eqs. \eqref{tauk} and \eqref{lab1}.

At $\tau=0$ the integral in Eq. \eqref{part1} can be evaluated explicitly
\beq
f^{(t)}(0,0)=-\frac{5}{2\pi}\mbox{Cl}_2\left(\frac{\pi}{3}\right)-\frac{1}{4}\log\frac{4}{3},\label{part4}
\eeq
where $\mbox{Cl}_2(x)$ is the Clausen function,
\beq
\mbox{Cl}_2\left(\frac{\pi}{3}\right)\approx 1.0149416064096536.
\eeq

To calculate the expansion of $f^{(t)}(\tau,0)$ for small $\tau>0$
we first calculate the derivative of Eq. \eqref{part1}
with respect to $\be$. It can be expressed in terms of
the complete elliptic integral of the first kind
\beq
\frac{df^{(t)}(\tau,0)}{d\be}=\frac{e^{2\be}(e^{4\be}-3)}{\pi\tanh2\be} K(1/k)\label{part5}
\eeq
and $k$ is given by Eq. \eqref{lab1}.

From Eqs. \eqref{tauk} and \eqref{lab1} we can express $\Delta\beta$ in Eq. \eqref{lab18} as series in $\tau$
 \begin{align}
\Delta\beta
=&\frac{\tau}{4}-\frac{\tau^2}{16}-\frac{11\tau^3}{192}+\frac{7\tau^4}{256}+\frac{39\tau^5}{1280}
 -\frac{205\tau^6}{12288}\nonumber\\
 -&\frac{1165 \tau^7}{57344} + \frac{767 \tau^8}{65536} + \frac{8887 \tau^9}{589824}+O(\tau^{10}).\label{part3}
 \end{align}

Expanding Eq. \eqref{part5} for small $\tau>0$ and integrating, we arrive at the high-temperature expansion
of the free energy \eqref{part1} at $H=0$
\beq
f^{(t)}(\tau,0)=\frac{m^2(\tau,0)}{8\pi}\log m^2(\tau,0)+A(\tau),\label{part6}
\eeq
\beq
m(\tau,0)=-C_\tau \tau a(\tau),\quad C_\tau=\frac{3^{3/4}}{\sqrt{2}}, \label{part7}
\eeq
Comparing this with Eq. \eqref{lab4}, we obtain series expansions for the functions $a(\tau)$ and $A(\tau)$

\begin{align} a(\tau)=
&1 -\frac{3 \tau^2}{16}+\frac{23 \tau^4}{256}-\frac{229 \tau^6}{4096}\nonumber\\
&+\frac{25819 \tau^8}{655360}+\mathcal O(\tau^{10}), \label{part8}
\end{align}
\begin{align}
&A(\tau)=-\frac{5}{2\pi}\mbox{Cl}_2\left(\frac{\pi}{3}\right)-\frac{1}{4}\log\frac{4}{3}+\frac{\tau}{2} -
\frac{\tau^3}{12} + \frac{3 \tau^5}{80}\nonumber\\
& - \frac{5 \tau^7}{224} + \frac{35 \tau^9}{2304}+\frac{\tau^2}{32}
\left[2-\frac{3\sqrt{3}(2+3\log12)}{\pi}\right]\nonumber\\
&+\frac{\tau^4}{256}\left[-7+\frac{3\sqrt{3}(5+9\log12)}{\pi}\right]+\nonumber\\
&+\frac{\tau^6}{8192}\left[\frac{410}{3}-\frac{9\sqrt{3}(28+55\log12)}{\pi}\right]+\nonumber\\
&+\frac{\tau^8}{65536}\left[-767+\frac{\sqrt{3}(1291+2682\log12)}{\pi}\right]+\ldots\label{part9}
\end{align}

Using the exact expression for a zero-field magnetization for $\tau<0$
\beq
M=(1-k^2)^{1/8}, \label{part10}
\eeq
we find the expression for the function $c(\tau)$ in Eq. \eqref{lab4}
\beq
c(\tau)=-\frac{M(\tau)}{C_H \tilde G_1(-C_\tau \tau a(\tau))^{1/8}},\label{part11}
\eeq
where the coefficient $\tilde G_1$ from Eq. \eqref{lab9} is known \cite{McCoyWu}
\beq
\tilde G_1=-2^{1/12}\,e^{-1/8}\,{\mathcal A}^{3/2}=-1.357838341706595\ldots\ ,\label{part12}
\eeq
and ${\mathcal  A}=1.282427\ldots$ is the Glaisher constant.

Using Eqs. \eqref{tauk} and \eqref{lab1}, we can also expand $k(\tau)$ in $\tau$
\beq
k(\tau)=1+2\tau+2\tau^2+\tau^3-\frac{\tau^5}{4}+\frac{\tau^7}{8}-\frac{5\tau^9}{64}+O(\tau^{11}).\label{part13}
\eeq
Combining Eqs. (\ref{part10}--\ref{part13}), we finally obtain
\beq
C_H=\frac{2^{11/48}e^{1/8}}{3^{3/32}{\mathcal  A}^{3/2}}=0.825075494181738\ldots,\label{part14}
\eeq
\begin{align}
&c(\tau)=1+\frac{\tau}{4} + \frac{15 \tau^2}{128} - \frac{9 \tau^3}{512} - \frac{1447 \tau^4}{32768}
+\frac{649 \tau^5}{131072}\nonumber\\
& + \frac{109293 \tau^6}{4194304} - \frac{29803 \tau^7}{16777216}
 - \frac{194751097 \tau^8}{10737418240}  +O(\tau^{9}).\label{part56}
\end{align}

\section{Susceptiblity\label{secsus}}

Our analysis greatly relies
on the availability of the high order perturbation theory calculations
of the magnetic susceptibility in the Ising model
on the triangular lattice \cite{CGNP11}.
As mentioned before, the first nontrivial
contribution from the CFT irrelevant operators in this case
comes only at the order
$\tau^6$ and this helps to obtain very accurate results for the
scaling function.
First, we start with our definition of susceptibility $\eqref{F-def}$ with the free energy given by $\eqref{lab2}$.
For simplicity, we consider the high-temperature regime $\tau>0$.
Substituting Eqs. \eqref{lab4}, \eqref{lab5a}, \eqref{lab6}, and \eqref{lab7}
into Eq. \eqref{lab2} and differentiating over $H$ twice one obtains,
\begin{align}
\chi(\tau)=&-\frac{2C_H^2G_2}{C_{\tau}^{7/4}\tau^{7/4}}
\frac{c(\tau)^2}{a(\tau)^{7/4}}-2(B(\tau)+\fsub(\tau))\nonumber\\
&+\frac{C_{\tau}\mu_h\tau a(\tau)b(\tau)}{2\pi}(1+2\log(c_\tau \tau a(\tau))),\label{sus1}
\end{align}
where $G_2$ is the first expansion coefficient in the scaling function $\ghigh(\xi)$ and given in Table \ref{tablef}.
It was evaluated with a very high accuracy in Ref. \cite{ONGP}.

The first term in Eq. \eqref{sus1} describes the contribution of the Aharony and Fisher scaling function \cite{AF80}
on a triangular lattice
\begin{align}
&F_{AF}(\tau)=\frac{c(\tau)^2}{a(\tau)^{7/4}}=k^{1/4}\biggl[1 + \frac{\tau^2}{2} - \frac{21 \tau^4}{256}\nonumber\\
& + \frac{85 \tau^6}{2048} -
\frac{8669 \tau^8}{327680} + \frac{49507 \tau^{10}}{2621440}+O(\tau^{12})\biggr] \label{sus1b}
\end{align}
and the coefficient $C_0$
\beq
C_0=-\frac{2C_H^2G_2}{C_{\tau}^{7/4}}=1.089549651052967\ldots \label{sus1c}
\eeq
coincides with $C_{0\pm}^{\mbox{\scriptsize tr}}$ from Ref. \cite{CGNP11}.

Now let us give the expression for the Ising susceptibility on a triangular lattice \cite{CGNP11}. It can be written
in the form
\beq
\chi_{\mbox{\scriptsize CGNP}}(\tau)=C_0\tau^{-7/4}F_{lat}(\tau)+B_{lat}(\tau),\label{sus2}
\eeq
\beq
B_{lat}(\tau)=\sum_{p=0}^\infty  b_{p,q}(\log\tau)^p\tau^{p^2}B_p(\tau),\label{sus3}
\eeq
where $B_p(\tau)$ are regular in $\tau$.
Let us note that the term with $p=3$ starts with the lowest power $\tau^9$ and its contribution to the scaling function
 is of the order $H^2\tau^9\sim \xi^2 \tau^{11+3/4}$.
 Functions $B_{p}(\tau)$, $p=0,1,2$ are given in Appendix C.3 of Ref. \cite{CGNP11} and we will not list them here.

 We have
 \beq
 F_{lat}(\tau)=F_{AF}(\tau)+\Delta F, \label{sus4}
 \eeq
 \beq
 \Delta F= k^{1/4}[c_6\tau^6+c_8\tau^8+c_{10}\tau^{10}+O(\tau^{12})], \label{sus5}
 \eeq
 \begin{align}
 c_6=-&0.1774838832948664\ldots,\nonumber\\
 c_8=-&\frac{1}{102400}-\frac{403}{400}c_6\nonumber\\
 =&0.1788052467945779\ldots,\nonumber\\
 c_{10}=-&0.1488704025260859\ldots \label{sus6}
 \end{align}
The term $\Delta F$ comes from  CFT irrelevant operators and contributes to the subleading
part $\fsub(\tau)$.

Finally, comparing Eqs. \eqref{sus1} and \eqref{sus2} we can determine $\mu_h$, $b(\tau)$, $B(\tau)$ and $\fsub(\tau)$.
We give them up to the order $\tau^5$, further terms' contributions to the free energy are less
than $10^{-15}$. We also restrict accuracy of the coefficients to $10^{-10}$ because all these functions
are multiplied by $H^2\sim \tau^{15/4}\approx 10^{-6}$ for values of $\tau\approx 0.025$.
\beq
\mu_h=-0.01047478006\ldots, \label{sus7}
\eeq
\begin{align}
b(\tau)= &1 + \frac{ \tau}{2} +
0.122460779\, \tau^2 + 9.228424771\, \tau^3\nonumber\\
 & -
 4.710909908\,\tau^4 +6.54812548\, \tau^5 +\ldots, \label{sus8}
\end{align}
\begin{align}
B(\tau)= &0.02478055826 + 0.02444328450 \tau\nonumber\\
 + &0.01102511536 \tau^2 - 0.001283871801 \tau^3\nonumber\\
  - &0.07567570347 \tau^4 - 0.03469707085 \tau^5+\ldots. \label{sus9}
\end{align}

Contributions to $\fsub(\tau)$ come from $\Delta F$ in \eqref{sus5} and terms
proportional to $(\log\tau)^2$ in Eq. \eqref{sus3}. We have
\begin{align}
&\fsub(\tau) =1.089549651\,\tau^{\frac{17}{4}}(1+\frac{\tau}{2}+0.8824449774\tau^2)\nonumber\\
&-0.0041507859(\log\tau)^2\tau^4(1+\frac{\tau}{2}-0.8222691101\tau^2). \label{sus10}
\end{align}

Let us also notice that for $\tau\approx0.025$ a contribution from the sub-leading term \eqref{sus10}
to the scaling function $G(\xi)$ is of the order $10^{-8}$.

Finally, we need to determine $e_h$ and $d(\tau)$ in Eq. \eqref{lab4}.
The value of the constant $e_h$ was accurately estimated in Ref. \cite{MBBD10}
\beq
e_h=0.00129(1).
\eeq
Since $e_h H^3 = e_h \xi^3\tau^{45/8} \approx  10^{-14}$ for $\tau=0.025$ and $\xi=0.2$,
we can expect that the linear term
in $d(\tau)$ will give a contribution $\sim 10^{-16}$ to the free energy.
We will neglect
it and simply choose
\beq
d(\tau)=1. \label{sus11}
\eeq
It should not affect the accuracy of the scaling function $G(\xi)$.

\section{CTMRG\label{secCTM}}

Corner transfer matrices (CTM) have proven to be an effective tool for numerical study of lattice systems
 in 2D \cite{Bax1968,Bax1978,Baxter2007}.
Nishino {\it et al.} \cite{Nishino97} introduced an improved iteration scheme for the original Baxter approach,
 now known as the corner transfer
matrix renormalization group (CTMRG). In this section we shall give a brief introduction to this method.

We first formulate the algorithm for a square lattice in the Interaction-Round-Face (IRF) formulation.
Although this is not essential for the algorithm, we assume that all spins take two values, $+1$ and $-1$.

First, we introduce the four-spin Boltzmann weight $w(a,b,c,d)$ as in Fig. \ref{W-picture}.

\begin{figure}[ht]
\setlength{\unitlength}{0.4mm}
\begin{picture}(200,60)(-60,20)
\multiput(50,30)(0,30){2}{\line(1,0){30}}
\multiput(50,30)(30,0){2}{\line(0,1){30}}
\put(40,30){\large $c$}\put(85,30){\large $d$}
\put(40,60){\large $a$}\put(85,60){\large $b$}
\put(-30,45){\large $w(a,b,c,d)=$}
\multiput(47.5,28)(30,0){2}{\wcircle{0.1}}
\multiput(47.5,58)(30,0){2}{\wcircle{0.1}}
\end{picture}
\caption{The four-spin weight $w(a,b,c,d)$.}
\label{W-picture}
\end{figure}

For a symmetric square lattice Ising model  we have
\beq
    w(a, b, c, d) = e^{\frac{\beta}{2}(ab + ac + bd + cd) + \frac{H}{4}(a +b+c+d)}. \label{ctm1}
\eeq
The factors $1/2$ and $1/4$ enter Eq. \eqref{ctm1} because each edge is shared by two plaquettes
and each vertex is shared by four plaquettes of the square lattice.

Now we define a half-row transfer matrix (HRTM) $F(a,b)_{i,j}$. For
the transfer matrix of the length $N+2$
we fix the  leftmost spins as $(a,b)$, the  rightmost spins as $(+1,+1)$ and combine
the remaining spins into multi-indices $i$ and $j$
\beq
i=\{i_1,\ldots,i_N\},\quad j=\{j_1,\ldots,j_N\} \label{ctm2}
\eeq
as shown in Fig. \ref{F-picture}.
\begin{figure}[ht]
\setlength{\unitlength}{0.3mm}
\begin{picture}(273,100)(-70,10)
\multiput(40,50)(0,30){2}{\line(1,0){120}}
\multiput(40,50)(30,0){5}{\line(0,1){30}}
\put(28,47){\large $a$}\put(97,20){\large{$i$}}
\put(28,77){\large $b$}\put(97,106){\large{$j$}}
\put(60,40){ $\underbrace{i_1\phantom{iiiiiiiiiiiiiiiii}}$}
\put(90,40){ $i_2$}\put(120,40){ $i_3$}\put(162,47){ $+1$}
\put(60,86){ $\overbrace{j_1\phantom{iiiiiiiiiiiiiiiii}}$}
\put(90,86){ $j_2$}\put(120,86){ $j_3$}\put(162,77){ $+1$}
\put(-50,60){\large $F(a,b)_{i,j}=$}
\multiput(37,48)(30,0){5}{\wcircle{0.08}}
\multiput(37,78)(30,0){5}{\wcircle{0.08}}
\end{picture}
\caption{The half-row transfer matrix $F(a,b)_{i,j}$}
\label{F-picture}
\end{figure}

The rightmost spins $(+1,+1)$  play the role of boundary
conditions and we will always fix boundary spins to $+1$ values. For 2D models out of criticality,
the choice of boundary spins should become irrelevant in the thermodynamic limit.
In general, we also need a half-column transfer matrix $G(a,b)$. However,
in the symmetric case $F(a,b)$ and $G(b,a)$
are related by the matrix transposition, see \eqref{ctm5}.

Now we define the corner transfer matrix (CTM) $A(a)_{i,j}$, see Fig. \ref{A-picture}.
Its rightmost and topmost indices
are fixed to $+1$, the bottom and left spins are combined into matrix indices $i$ and $j$ except the
corner spin $a$.
\begin{figure}[ht]
\setlength{\unitlength}{0.3mm}
\begin{picture}(273,180)(-50,10)
\multiput(70,80)(30,0){3}{\circle*{6}}
\multiput(70,110)(30,0){3}{\circle*{6}}
\multiput(70,140)(30,0){3}{\circle*{6}}
\multiput(40,50)(0,30){5}{\line(1,0){120}}
\multiput(40,50)(30,0){5}{\line(0,1){120}}
\put(28,43){\large $a$}\put(97,15){\large{$i$}}
\put(7,106){\large{$j$}}
\put(61,38){ $\underbrace{i_1\phantom{iiiiiiiiiiiiiiii,}}$}
\put(91,38){ $i_2$}\put(121,38){ $i_3$}
\put(26,77){$ j_1$}
\put(15,105){$ \left\{\phantom{\begin{array}{l}I\\I\\I\\I\\I\\I\end{array}}\right.$}
\put(21,107){ $j_2$}\put(21,137){ $j_3$}
\multiput(34,176)(30,0){4}{ $+1$}
\multiput(162,48)(0,30){4}{ $+1$}
\put(163,175){$+1$}
\multiput(37,48)(30,0){5}{\wcircle{0.08}}
\multiput(37,168)(30,0){5}{\wcircle{0.08}}
\multiput(37,78)(0,30){3}{\wcircle{0.08}}
\multiput(157,78)(0,30){3}{\wcircle{0.08}}
\end{picture}
\caption{The corner transfer matrix $A(a)_{i,j}$}
\label{A-picture}
\end{figure}

We also sum over all internal spins shown by black circles. Here and below
we follow the notation, that the spins denoted by white circles (or
rectangles for multi-indices) are fixed, while the spins denoted by
black circles/rectangles are summed over.

Notice that the $2^N\times 2^N$ corner transfer matrix $A(a)$ is
nothing but the partition function of the model on the
$(N+2)\times(N+2)$ square lattice with fixed boundary conditions.

The main idea of the CTMRG method is to calculate $A(a)$ recursively
and truncate it at each step
to physically relevant degrees of freedom.
The core of the algorithm
relies on the insight that the eigenvalues
of the transfer matrix decay exponentially fast in the off-critical regime,
so the vast majority of the information
about the CTMs is contained in a finite set of dominant eigenvalues.
At each step, the iteration algorithm doubles
the size of $A$ and $F$, but since the majority
of information is contained in the largest eigenvalues,
the smaller half of eigenvalues can be discarded
with minimal information loss while reducing the size
of the matrix to its original size.

We can summarize the algorithm by the following steps.

1. We start the algorithm with the initialization of matrices $A$ and $F$. It is easy to calculate them for
small-size lattices. We could even initialize them as random
matrices with positive entries, this does not much affect the
convergence of the algorithm.

2. We update the  CTM $A(a)_{i,j}$ to the expanded CTM $A'(a)_{I,L}$
of a double size as shown in Fig. \ref{R-picture}.
\begin{figure}[ht]
\setlength{\unitlength}{0.3mm}
\begin{picture}(273,170)(-50,-10)
\multiput(20,50)(0,90){2}{\line(1,0){120}}
\multiput(50,20)(90,0){2}{\line(0,1){120}}
\put(20,20){\line(1,0){120}}
\put(20,20){\line(0,1){120}}
\multiput(17,18)(0,30){2}{\wcircle{0.1}}
\put(47,18){\wcircle{0.1}}
\put(50,50){\circle*{7}}
\put(55,16){
\tikz \draw[color=black, fill=white]
 (0,0) rectangle (2.3cm,1.5ex);}
\put(55,46){
\tikz \draw[color=black, fill=black](0,0) rectangle (2.3cm,1.5ex);}
\put(43,58){
\tikz \draw[color=black, fill=black](0,0) rectangle (0.18cm,17ex);}
\put(13,58){
\tikz \draw[color=black, fill=white](0,0) rectangle (0.18cm,17ex);}
\put(56,90){\large$ k$}\put(26,90){\large$ l$}\put(18,153){ $G(d,b)$}
\put(90,58){\large$ j$}\put(90,28){\large$ i$}\put(150,32){ $F(c,d)$}
\put(80,90){\large $ A(d)$}
\put(40,40){$d$}\put(12,12){$a$}
\put(12,39.5){$b$}\put(41,12){$c$}\put(28,28){\large $\mathbf w$}
\put(48,8){$\underbrace{\phantom{iiiiiiiiiiiiiiiiiiiiiiii}}$}
\put(0,87){$ \left\{\phantom{\begin{array}{l}I\\I\\I\\I\\I_I\\,\\,\\,\end{array}}\right.$}
\put(90,-10){$I$}\put(-10,90){$L$}
\put(148,35){\vector(-1,0){20}}\put(35,148){\vector(0,-1){20}}
\end{picture}
\caption{The updated  matrix $A'(a)_{I,L}$}
\label{R-picture}
\end{figure}

This can be  written as
\beq
A'(a)_{I,L}=\sum_{d,j,k}w(b,d,a,c)F(c,d)_{ij}A(d)_{j,k}G(d,b)_{k,l}\label{ctm3}
\eeq
with $I=\{c,i\}$, $L=\{b,l\}$.

If the weights $w(a,b,c,d)$ satisfy the symmetry
\beq
w(a,b,c,d)=w(d,b,c,a), \label{ctm4}
\eeq
then the CTM $A(a)$ will be a symmetric matrix and
the half-column matrix transfer matrix $G(a,b)$ is a transposition of $F(b,a)$
\beq
G(a,b)_{i,j}=F(b,a)_{j,i}. \label{ctm5}
\eeq

We update the HRTM $F(a,b)$ by simply adding another weight $w$ on the left.
After several iterations we arrive at matrices of reasonable sizes
the computer can still handle reasonably well, say $64\times 64$.

3. Now we diagonalize $A'(a)$ for $a=\pm1$
\beq
A'(a)=U(a)A'_d(a)U^{-1}(a). \label{ctm6}
\eeq
For a real magnetic field, the matrix $A'(a)$ is a real symmetric matrix.
Therefore, the matrix $U(a)$ can be chosen orthogonal.
The columns of $U(a)$ are the eigenvectors of $A'(a)$.

In this paper we investigate the Ising model
in a complex magnetic field $H$.
The CTM in this case will be a complex symmetric matrix.
Its eigenvalues $A'_d(a)$ in Eq. \eqref{ctm6} will, in general, be complex.
We will order them by their absolute value. The numerical calculations
clearly show that all these eigenvalues are non-degenerate (as it would
reasonably be expected). Note also,
that the matrix $U(a)$ in Eq. \eqref{ctm6}
in this case will no longer be orthogonal or unitary.

As explained in Sec. \ref{secscal},
for  a fixed value of the temperature-like parameter $\tau$
 the free energy of the model
is an analytic function of the magnetic field $H$
in the cut plane with the branch cuts associated
with Yang-Lee edge singularity.
We have carefully investigated
a stability of our calculations with respect to small imaginary variations
of the magnetic field.
The results were always consistent and independent of the path chosen
to approach complex values of $H$ starting from the
purely real ones, as long as the path does not cross the Yang-Lee branch cuts.

4. If the dimension of the original CTM $A(a)$ is $N$, then the dimension of the updated matrix
$A'(a)$ will be $2N$. We now select first $N$  eigenvalues of $A'_d(a)$ and form $2N\times N$
matrix from the eigenvectors
\beq
V(a)_{I,j}=U(a)_{I,j},\quad I=1,\ldots,2N,\quad j=1,\ldots N.\label{ctm7}
\eeq
We notice that the first index of $V(a)$ still has the tensor structure $I=\{c,i\}$ inherited from Eq. 
\eqref{ctm3}.

\begin{figure}[ht]
\setlength{\unitlength}{0.3mm}
\begin{picture}(273,117)(-90,-7)
\multiput(50,10)(0,90){2}{\line(1,0){90}}
\multiput(20,40)(0,30){2}{\line(1,0){120}}
\multiput(20,40)(30,0){2}{\line(0,1){30}}
\put(140,10){\line(0,1){90}}
\put(20,40){\line(1,-1){30}}
\put(20,70){\line(1,1){30}}
\put(110,-3){$k$}\put(110,105){$l$}
\put(110,47){$i$}\put(110,79){$j$}
\multiput(17,38)(0,30){2}{\wcircle{0.1}}
\put(50,40){\circle*{7}}
\put(50,70){\circle*{7}}
\put(60,36){\tikz \draw[color=black, fill=black](0,0) rectangle (2.3cm,1.5ex);}
\put(60,66){\tikz \draw[color=black, fill=black](0,0) rectangle (2.3cm,1.5ex);}
\put(60,6){\tikz \draw[color=black, fill=white](0,0) rectangle (2.3cm,1.5ex);}
\put(60,96){\tikz \draw[color=black, fill=white](0,0) rectangle (2.3cm,1.5ex);}
\put(-75,52){ \large$F_n(a,b)_{k,l}=$}
\put(55,52){ $F(c,d)$}\put(60,20){$V^T(a)$}\put(60,80){$V(b)$}
\put(40,60){$d$}\put(12,32){$a$}
\put(12,59.5){$b$}\put(41,32){$c$}\put(28,48){\large $\mathbf w$}
\end{picture}
\caption{The updated  matrix $F_n(a,b)$}
\label{F-update}
\end{figure}
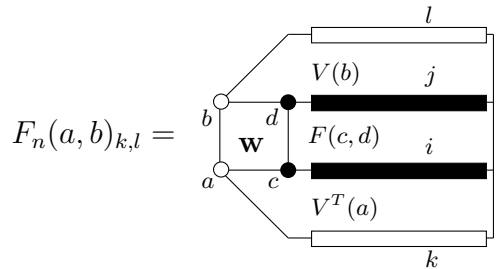

Now we form new $N\times N$ CTM $A_n(a)$ and HRTM $F_n(a,b)$
\beq
A_n(a)_{i,j} = \delta_{ij}A'_d(a)_{i,i}, \quad i,j=1,\ldots, N \label{ctm8}
\eeq
\begin{align}
&F_n(a,b)_{k,l}=\sum_{c,d,i,j}w(b,d,a,c)\nonumber\\
&\times V^T(a)_{k,\{c,i\}}F(c,d)_{i,j}V(b)_{\{d,j\},l} \label{ctm9}
\end{align}
as shown in Fig. \ref{F-update}.

5. Then we go back to step 2 and repeat iterations until the process converges.
We may also increase the size $N$ of the transfer matrices to get a better convergence.
We used Intel Fortran with the maximum value $N=170$ to achieve the error comparable with the machine
accuracy $10^{-16}$ for a range of temperatures and magnetic fields near the Yang-Lee singularity.

It is possible to achieve a much higher accuracy with larger values of $N$ performing calculations
with quad precision. For example, in the high-temperature regime, our numerical free energy
matched the Onsager's result with $10^{-25}$ accuracy. However, this will not improve the accuracy
of the scaling function because we need to know series in $\tau$ to higher orders which are determined
by irrelevant operators and very hard to control.

\section{Partition function per site\label{secpart}}

Once we calculated the CTM $A$ and the HRTM $F$, we can calculate the partition function per cite following
Baxter's variational arguments \cite{Bax1968, Bax1978}. We shall refer the interested reader to the original
Baxter's papers and first give the result for a square lattice. The partition function per site $\kappa$
is given by
\beq
    \kappa = \frac{r_1r_4}{r_2 r_3}\label{pps1}
\eeq
with each term on the right given explicitly by
\begin{align}
    r_1 = \sum_a &\Tr[A(a)^4]\nonumber \\
    r_2 = \sum_{a, b} &\Tr[A(a)^2 F(a, b)A(b)^2F(b, a)]\nonumber\\
    r_3 = \sum_{a, b} &\Tr[A(a)^2 F(b,a)^TA(b)^2F(a,b)^T]\nonumber \\
    r_4 = \sum_{a, b, c, d} &w(a, b,c,d)\Tr\left[A(a) F(a, c)A(c)F(d,c)^T\right.\nonumber \\
    &\left.A(d)F(d,b)A(b)F(a,b)^T\right].\label{pps2}
\end{align}
For the symmetric Ising model, $A(a)$ is symmetric and
\beq
F(a,b)=F(b,a)^T. \label{pps2a}
\eeq

Now let us discuss the case of the symmetric Ising model on a triangular lattice. We can reduce the CTMRG
algorithm to the square lattice one by choosing the weight in the form
\beq
w(a, b, c, d) = e^{\frac{\beta}{2}(ab + ac + bd + cd+2ad) + \frac{H}{6}(2a +b+c+2d)}, \label{pps3}
\eeq
see Fig. \ref{tri-lat}.

\begin{figure}[ht]
\setlength{\unitlength}{0.4mm}
\begin{picture}(200,60)(-60,20)
\multiput(50,30)(15,30){2}{\line(1,0){30}}
\multiput(50,30)(30,0){2}{\line(1,2){15}}
\put(65,60){\line(1,-2){15}}
\put(40,30){\large $c$}\put(85,30){\large $d$}
\put(55,60){\large $a$}\put(100,60){\large $b$}
\put(-30,45){\large $w(a,b,c,d)=$}
\multiput(47.5,28)(30,0){2}{\wcircle{0.1}}
\multiput(62.5,58)(30,0){2}{\wcircle{0.1}}
\end{picture}
\caption{The four-spin weight $w(a,b,c,d)$ for a triangular lattice.}
\label{tri-lat}
\end{figure}

The square lattice CTMRG algorithm will still work since  the weight \eqref{pps3} satisfies the property \eqref{ctm4}
and the matrix $A(a)$ is symmetric. However, the matrix $F(a,b)$ no longer
satisfies the property \eqref{pps2a} and we need to use $F^T$.
We also need to modify a calculation of the partition function per site.

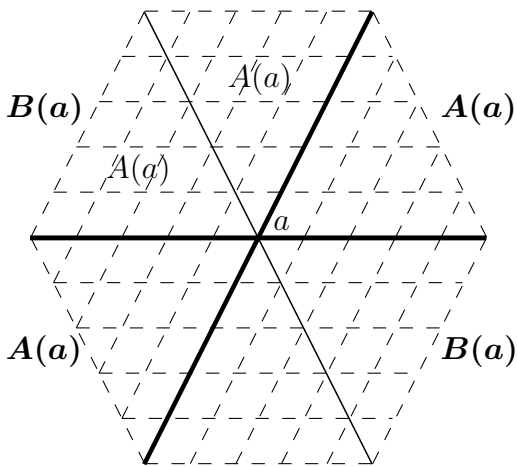
\begin{figure}[ht]
\setlength{\unitlength}{0.4mm}
\begin{picture}(200,175)(-30,5)
\put(0,90){\linethickness{0.06cm}\line(1,0){150}}
\put(75,90){\linethickness{0.06cm}\line(1,2){37.5}}
\put(75,90){\linethickness{0.02cm}\line(-1,2){37.5}}
\multiput(7.5,0)(7.5,15){5}{
\put(0,105){\tikz \draw [dash pattern={on 5pt off 5pt }] (0,0) -- (3,0);}}
\put(87,105){\tikz \draw [dash pattern={on 5pt off 5pt }] (0,0) -- (2.2,0);}
\put(95,120){\tikz \draw [dash pattern={on 5pt off 5pt }] (0,0) -- (1.6,0);}
\put(102,135){\tikz \draw [dash pattern={on 5pt off 5pt }] (0,0) -- (1,0);}
\put(109,150){\tikz \draw [dash pattern={on 5pt off 5pt }] (0,0) -- (0.4,0);}
\multiput(7.5,0)(7.5,-15){5}{
\put(0,75){\tikz \draw [dash pattern={on 5pt off 5pt }] (0,0) -- (3,0);}}
\put(87,75){\tikz \draw [dash pattern={on 5pt off 5pt }] (0,0) -- (2.2,0);}
\put(95,60){\tikz \draw [dash pattern={on 5pt off 5pt }] (0,0) -- (1.6,0);}
\put(102,45){\tikz \draw [dash pattern={on 5pt off 5pt }] (0,0) -- (1,0);}
\put(109,30){\tikz \draw [dash pattern={on 5pt off 5pt }] (0,0) -- (0.4,0);}
\put(112.5,90){\tikz \draw [dash pattern={on 5pt off 5pt }] (0,0) -- (-1.5,3);}
\put(0,15){\tikz \draw [dash pattern={on 5pt off 5pt }] (0,0) -- (-1.5,3);}
\multiput(2.5,0)(15,0){5}{
\put(-2.5,90){\tikz \draw [dash pattern={on 5pt off 5pt }] (0,0) -- (1.5,3);}}
\put(30,30){\tikz \draw [dash pattern={on 5pt off 5pt }] (0,0) -- (1.17,2.34);}
\put(22.5,45){\tikz \draw [dash pattern={on 5pt off 5pt }] (0,0) -- (0.9,1.8);}
\put(15,60){\tikz \draw [dash pattern={on 5pt off 5pt }] (0,0) -- (0.6,1.2);}
\put(7.5,75){\tikz \draw [dash pattern={on 5pt off 5pt }] (0,0) -- (0.3,0.6);}
\put(90,90){\tikz \draw [dash pattern={on 5pt off 5pt }] (0,0) -- (1.17,2.34);}
\put(105,90){\tikz \draw [dash pattern={on 5pt off 5pt }] (0,0) -- (0.9,1.8);}
\put(120,90){\tikz \draw [dash pattern={on 5pt off 5pt }] (0,0) -- (0.6,1.2);}
\put(135,90){\tikz \draw [dash pattern={on 5pt off 5pt }] (0,0) -- (0.3,0.6);}
\multiput(40,-75)(15,0){6}{
\put(-2.5,90){\tikz \draw [dash pattern={on 5pt off 5pt }] (0,0) -- (1.5,3);}}
\put(0,90){\linethickness{0.06cm}\line(1,0){150}}
\put(75,90){\linethickness{0.02cm}\line(1,-2){37.5}}
\put(75,90){\linethickness{0.06cm}\line(-1,-2){37.5}}
\put(135,130){\large$\boldsymbol{A(a)}$}\put(80,92.5){\large $a$}
\put(-8,130){\large$\boldsymbol{B(a)}$}\put(-8,50){\large$\boldsymbol{A(a)}$}
\put(135,50){\large$\boldsymbol{B(a)}$}
\put(65,140){\large $A(a)$}\put(25,110){\large $A(a)$}
\end{picture}
\caption{CTMs for a triangular lattice.}
\label{CTMtri-lat}
\end{figure}

First, we draw a triangular lattice in the form of four parts separated by bold lines where each dashed rectangle
is identified with the weight \eqref{pps3}. We can identify this with a square lattice
with two transfer matrices $A(a)$ and $B(a)$ assigned to different quadrants, see Fig. \ref{CTMtri-lat}.

It is clear from Fig. \ref{CTMtri-lat} that the $B(a)$ is equal to $A^2(a)$.
Therefore, they are both symmetric and
diagonalized by the same transformation.

A (nonnormalized) partition function $r_1$ becomes
\beq
r'_1=\sum_a \Tr[A(a)B(a)A(a)B(a)]=\sum_a\Tr[A^6(a)].\label{pps4}
\eeq

Repeating Baxter's arguments for the partition function per site, we come to the expression
\beq
    \kappa_t = \frac{r'_1r'_4}{r'_2 r'_3}\label{pps5}
\eeq
with
\begin{align*}
    r_2' = &\sum_{a, b} \Tr[A(a)^3 F(a, b)A(b)^3F(b, a)]\\
    r_3' = &\sum_{a, b} \Tr[A(a)^3 F(b,a)^TA(b)^3F(a,b)^T]\\
    r_4' = &\sum_{a, b, c, d} w(a, b,c,d)\Tr\left[A(a)^2 F(a, c)A(c)F(d,c)^T\right.\nonumber \\
    &\qquad\left.A(d)^2F(d,b)A(b)F(a,b)^T\right],\label{pps6}
\end{align*}
where $w(a,b,c,d)$ is given by Eq. \eqref{pps3}. Note also, that the HRTM
$F$ no longer satisfies Eq. \eqref{pps2a}.

We can now apply the square lattice CTM algorithm to calculate the
transfer matrices $A(a)$ and $F(a,b)$ and then
calculate the partition function per site using Eq. \eqref{pps5}. The free energy per site is simply given by
\beq
\fctm(\tau,H)=-\log k(\tau,H).\label{pps7}
\eeq

\section{Numerical results\label{secnum}}

In this section, we describe numerical calculations used to
generate and analyse the data for
the scaling function in the region close to the Yang-Lee
singularity. We will consider the high-temperature regime $m<0$ with a
complex magnetic field $H$. As explained before, the scaling function
$\ghigh(\xi)$ has two branch cuts $(-i\infty,-i\xi_0)$ and
$(i\xi_0,+i\infty)$ with the value $\xi_0$ estimated in Ref. \cite{FZ03}
and quoted here in Eq. \eqref{lab14}.

Following  Ref. \cite{FZ03}, let us introduce the function $\Phiimh(z)$ (and
a simply related function $\gly(z)$)
via
an analytic continuation of $\ghigh(\xi)$
to purely imaginary $\xi$. Here we define
\beq
\Phiimh(z) =z^2\gly(z)=z^2\ghigh(-iz^{-15/8}), \label{num1}
\eeq
noting that our variable $z$ is equal to $(-y)$ in the original
definition of this function in Eq. (3.38b) of Ref. \cite{FZ03}.
In terms of the nonlinear scaling variables \eqref{lab4} it reads
\beq\label{zvardef}
z=(i\xi)^{-8/15}=-m\,{(i h)^{-8/15}}\,.
\eeq
The functions $\Phiimh(z)$ and $\gly(z)$ have a branch cut on the real axis
for $0<z<Z_0$, where \cite{FZ03} [cf. \eqref{lab14}]
\beq Z_0=1/\xi_0^{8/15}\approx 2.4295\,.\label{Z0}
\eeq
Below, for series expansions
it will be convenient to use the variable
\beq\label{x-def}
v=(z-Z_0)=\frac{4}{15}Z_0^{19/4} u+\frac{38}{225}Z_0^{17/2} u^2+\cdots,
\eeq
where $\xi=-iz^{-15/8}$ and
\beq\label{u-def}
u=\xi^2+\xi_0^2=\frac{15}{4}Z_0^{-19/4}\, v -\frac{285}{32}
\,Z_0^{-23/4}\, v^2+\cdots\,.
\eeq

The scaling function $\gly(z)$ has a concise interpretation in terms of 2D
Euclidean quantum field theory. Namely, it coincides with the vacuum
energy density of the Ising field theory (IFT) in the vicinity of
the Yang-Lee singularity. The IFT is defined as the $c=1/2$
CFT perturbed by the energy and spin operators \cite{FZ03},
\beq\label{aift}
\mathcal{A}_\text{IFT}=\mathcal{A}_{(c=1/2)}+\frac{m}{2\pi}\int \epsilon(x)\,
d^2 x+h\,\int \sigma(x)\,d^2 x\,,
\eeq
where $\mathcal{A}_{(c=1/2)}$ stands for the action of the $c=1/2$
CFT of free massless Majorana fermions, $\sigma(x)$  and $\epsilon(x)$
are primary fields of conformal dimensions $(\frac{1}{16},\frac{1}{16})$
and $(\frac{1}{2},\frac{1}{2})$.
Their normalization is fixed by the usual CFT convention,
\beq
x^2 \langle\epsilon(x)\epsilon(0)\rangle\to1\,,\>\>
x^{1/4} \langle\sigma(x)\sigma(0)\rangle\to1\,,\   \mbox{as}\ \  x\to0\,.
\eeq
The coupling constants $m$ and $h$, appearing in Eq. \eqref{aift}, are the nonlinear scaling variables,
related to the lattice model parameter $\tau$ and $H$ via Eq. \eqref{lab4}. Remarkably, the same field theory can also
be defined [up to a constant shift of
the vacuum energy density, see Eq. \eqref{gterm} below] as a model of perturbed minimal conformal field
theory ${\cal M}_{2/5}$ (YLCFT)
\beq\label{aeff}
\mathcal{A}_\text{eff} =\mathcal{A}_\text{YLCFT}
+ \lambda\int\phi(x) \, d^2 x +\sum_{i} a_i\int O_i(x)\,d^2 x\,,
\eeq
involving an
infinite tower of irrelevant operators $O_i(x)$,
with dimensions $(\Delta_i,\overline{\Delta}_i)$ such that
$\Delta_i=\overline{\Delta}_i$ (see Ref. \cite{XuZ22} for the details).
In this work, we take into account the first few ``least irrelevant'' operators
\bea
\sum_{i} \,&a_i\int O_i(x)\,d^2 x=
+ \frac{\alpha}{\pi^2} \,\int T{\bar T}(x) \, d^2 x\nonumber\\
& +{\frac{\beta}{2\pi}}\, \int \Xi(x) \, d^2 x
+\frac{\gamma}{2\pi}\,\int \Xi_6(x) \, d^2 x
\label{aeff2}\\[.2cm]
& +\frac{\alpha_5}{\pi^2}\,\int (T{\bar T})^3(x) d^2 x
+\frac{\delta}{2\pi}\,\int \Xi_8(x) \, d^2 x
\nonumber\\[.2cm]
&+\mbox{higher irrelevant operators}\,.\nonumber
\end{align}
The action \eqref{aeff}
involves the ${\cal M}_{2/5}$
relevant primary operator $\phi$ with the conformal dimensions
$(-\tfrac{1}{5},-\tfrac{1}{5})$ and the irrelevant operators
$T\overline{T}$ and $(T\overline{T})^3$
with the dimensions $(2,2)$ and $(6,6)$ respectively;
the operators $\Xi$, $\Xi_6$ and $\Xi_8$ with the dimensions
$(\tfrac{19}{5},\tfrac{19}{5})$, $(\tfrac{29}{5},\tfrac{29}{5})$ and
$(\tfrac{39}{5},\tfrac{39}{5})$.
The latter are the level four, level six and level eight  descendants of the
primary field $\phi$.
We will disregard all operators
with the mass dimension greater than $78/5$, since their contribution
is too small to be properly accounted for by our
numerical data.
Here we assume the same definitions and normalizations of these
operators as in Refs. \cite{XuZ22,XuZ23}.

The coupling constant $\lambda$ carries the mass dimension
$[\lambda] \sim  [\mbox{mass}]^{12/5}$, while the constants
$\alpha$, $\beta$, $\gamma$ and $\delta$
have negative mass
dimensions, $\alpha\sim[\mbox{mass}]^{-2}$  and
$\beta\sim[\mbox{mass}]^{-28/5}$, $\gamma\sim[\mbox{mass}]^{-48/5}$, and
$\delta\sim[\mbox{mass}]^{-68/5}$.
In IFT these coupling constants depend on the scaling parameter
$\xi$ or $z$,
and admit convergent series expansion in powers of $v=(z-Z_0)$, in particular,
\begin{align}
&\lambda(z)=\sum_{k=1}^\infty\hat\lambda_k\,(z-Z_0)^k=\hat\lambda_1 v
  +\hat\lambda_2 v^2+
\cdots\,,\nonumber\\[.2cm]
&\alpha(z) = \sum_{k=0}^\infty\hat\alpha_k\,(z-Z_0)^k=
\hat\alpha_0 + \hat\alpha_1 v
+ \cdots\nonumber\,,\\
&\beta(z) = \sum_{k=0}^\infty\hat\beta_k\,(z-Z_0)^k
=\hat\beta_0 + \hat\beta_1 v 
+\cdots\label{alphaz}\,,\\
&\gamma(z) = \sum_{k=0}^\infty\hat\gamma_k\,(z-Z_0)^k
=\hat\gamma_0 + \hat\gamma_1 v 
+\cdots\nonumber\,,\\
&\delta(z) = \sum_{k=0}^\infty\hat\delta_k\,(z-Z_0)^k
=\hat\delta_0 + \hat\delta_1 v 
+\cdots\nonumber\,.
\end{align}

The QFT defined by the first two terms in the action
\eqref{aeff} is called the
Yang-Lee QFT. It is an integrable QFT containing one scalar particle
with the mass
\beq
\myl=\cyl\, \lambda^{5/12}\,,\label{mdef}
\eeq
where \cite{AlZam95}
\beq
\cyl=\frac{2^{19/12} \sqrt{\pi }}{5^{5/16}}
   \frac{\left(\Gamma
   \left(\frac{3}{5}\right) \Gamma
   \left(\frac{4}{5}\right)\right)^{
   5/12}}{\Gamma
   \left(\frac{2}{3}\right) \Gamma
   \left(\frac{5}{6}\right)}.
\eeq

With these notations 
the vacuum energy density of IFT defined by the action \eqref{aeff} reads
\bea
F_{\text{sing}}=
&\fyl \myl^2+ \sum_{i}\,c_i\, a_i\, \myl^{2\Delta_i}\nonumber \\[.2cm]
& + \sum_{i,j}\,c_{ij}\, a_i \,a_j\,\myl^{2\Delta_i+2\Delta_j-2}+\cdots\label{fsing-gen}
\end{align}
where $\fyl=-\sqrt{3}/12$ and numerical
constants $c_i,\,c_{ij},\ldots$ are expressed through
the vacuum expectation values of the products of the operators $O_i(x)$.
Remarkably, the contribution of the operator $T\overline{T}(x)$,
generated by the first term in Eq. \eqref{aeff2}, can be calculated to all
orders \cite{Smirnov:2016lqw}
in the coupling constant $\alpha$ in Eq. \eqref{aeff2}. The result is
\cite{XuZ22}
\beq\label{Fsing}
F_{\text{sing}}=\frac{F_{\text{sing}}^{(0)}}
{1+\alpha\,F_{\text{sing}}^{(0)}}\,,
\eeq
where $F_{\text{sing}}^{(0)}$ is defined by Eq. \eqref{fsing-gen} with $\alpha=0$.

Next note,
that the energy density $F_{\text{sing}}$ defined above is, of course,
a function of the scaling variable $z$, since the coupling constants
\eqref{alphaz} are the series in $z$.
The scaling function $\gly(z)$
introduced in Eq. \eqref{num1} can now be expressed as
\beq
\gly(z)=g(z)+F_{\text{sing}}(z)\,,\label{gterm}
\eeq
where the function $g(z)$ is analytic near $z=Z_0$,
\beq
g(z)=\sum_{k=0}^\infty \hat g_k\,(z-Z_0)^k= \hat g_0+\hat
g_1\,(z-Z_0)+\cdots\,, \label{gz}
\eeq
it can be viewed as an ``induced cosmological term'' \cite{XuZ22}
for the IFT \eqref{aeff}. Note, that $\hat g_0=\gly(Z_0)$.

The formula \eqref{gterm} allows one to parametrize the scaling function
$\gly(z)$ by the coefficients of the series expansions \eqref{alphaz}
and \eqref{gz}. Our aim is to determine the leading coefficients of
these series by fitting the results of the numerical CTM calculations of
the scaling function. Obviously, we need to reasonably truncate the
series \eqref{alphaz} so that the contribution of dropped terms
does not exceed the error of the numerical data used for the fit
[looking ahead, we will drop all terms, contributing powers higher
than $O[(z-Z_0)^8]$ into the series \eqref{gterm}].

Next, note that irrespective of such truncation, the series \eqref{alphaz}
contain an overdetermined set of coefficients, some of which cannot be
found solely from the knowledge of the scaling function.
To illustrate this point consider a few first terms in the
expansion $F^{(0)}_{\text{sing}}(z)$, given by Eq. \eqref{fsing-gen} with
$\alpha=0$,
\bea
F_{\text{sing}}^{(0)}=&\fyl \myl^2+c_{\Xi}\,\beta
\myl^{38/5}+c_{\Xi_6}\,\gamma \myl^{58/5}\nonumber\\[.2cm]
&+ O(\myl^{66/5})\label{3term}\,.
\end{align}
where the constants $c_{\Xi}$ and $c_{\Xi_6}$ are expressed via the vacuum
expectation values of the corresponding operators, appearing
in Eq. \eqref{aeff2}, see Eq. (2.5) in Ref. \cite{XuZ22}.
Now, make the following
transformations.
First, let us absorb the coefficients
$\hat\gamma_k$, $k=0,1,2,\ldots$ by redefining
$\hat\lambda_{k+5}$ in the first term of Eq. \eqref{3term}. Next, since
the coefficients $\hat\lambda_k$ also enter the second term via
$\myl$ one can compensate their change there by redefining $\hat\beta_k$,
$k=4,5,\ldots$. Therefore, the third term in Eq. \eqref{3term} is ``redundant''
as it could be completely absorbed into the first two.
Performing a similar analysis and omitting other
redundant terms one can write
$F_{\text{sing}}^{(0)}$ as
\bea
F_{\text{sing}}^{(0)}=&\fyl \myl^2+c_{\Xi}\,\beta
\myl^{38/5}+c_{\Xi\Xi}\,\beta^2 \myl^{66/5}\nonumber\\[.2cm]
&+c_{\Xi_8}\,\delta
\myl^{78/5}+ O(\myl^{88/5})\label{exp1}\,.
\end{align}
We stress that the above reduction of the ``redundant''
terms does not mean that they do not contribute.
We just state, that to write a series expansion
for $F_{\text{sing}}^{(0)}$ suitable for the fitting procedure, the
contribution of these terms
will be accounted for via a redefinition of the higher
coefficients entering the remaining terms in Eq. \eqref{exp1}. In
particular, the modified
coefficients $\hat\lambda_{k+5}$ and $\hat\beta_{k+4}$, with $k\ge0$,
will contain additions proportional to $\hat\gamma_k$. Moreover, the
``induced cosmological term'' \eqref{gz} will have an addition
proportional to $\sim(z-Z_0)^5\, \alpha_5(z)$ coming from
the contribution of the $(T\overline{T})^3$ term, contained in Eq. \eqref{aeff2}.

It is instructive to consider the series expansion of
the scaling function \eqref{gterm}.
A simple (but a bit tedious) analysis of the possible powers of the
variable $v=(z-Z_0)$ (to within the order of $O(v^8)$) leads to the following
expression
\beq\label{num2}
\gly(z)=b_0\,v^{19/6}+\sum_{\substack{0\le5k/6+l\le8\\0\le k\le5}}  a_{k,l}\,v^{5k/6+l}.
\eeq
It is worth noting that the coefficients
in this formula will not be all independent,
since, to within the same order in $v$,
the expression
\eqref{gterm} contains a lesser number of different coefficients than
Eq. \eqref{num2}. Note also, that using Eq. \eqref{x-def} one could easily rewrite
Eq. \eqref{num2} as a series in the variable $u$ defined in Eq. \eqref{u-def}.
The reason we prefer the variable $v$ is because the expansion
coefficients $a_{k,l}$ in Eq. \eqref{num2}
do not grow as fast as those for the corresponding expansion in $u$.
Moreover, we could compare our results for the first coefficients
$a_{k,l}$ with those obtained in Ref. \cite{FZ03}.

Since only an approximate location of the singularity
$Z_0$ is known, one must use a nonlinear fit to simultaneously
determine the value of $Z_0$ and coefficients $a_{k,l}$ and $b_0$ in
Eq. \eqref{num2}. Given this, we used the following iteration
procedure.

Starting with
some value of $Z_0$, initially given by Eq. \eqref{Z0}, the
function $\Phiimh(z)$ was calculated from Eqs. \eqref{num1} and \eqref{lab15}
using the CTM method for about thirty points in the
vicinity of $Z_0$.
Then we used the {\it NonlinearModelFit} package from {\it
  Mathematica} to determine the coefficients in Eq. \eqref{num2} and
a new value of $Z_0$.
Subsequently, the same procedure was repeated with an
updated value of $Z_0$ on each step.

Quite remarkably, the above iterations quickly converge to a rather
accurate value of $Z_0$,
\beq
Z_0^{\mbox{\scriptsize init}}=2.429169172\,, \label{num3a}
\eeq
which was used as an initial value for subsequent calculations.
Alternatively, we have also used the nonlinear formula \eqref{Fsing}
to simultaneously determine $Z_0$ and the coefficients \eqref{alphaz},
\eqref{gz}, but this led to the same result \eqref{num3a}.

Next, we have generated approximately 2000 points in the neighbourhood
of $Z_0^{\mbox{\scriptsize init}}$,
\beq
z=Z_0^{\mbox{\scriptsize init}}+ (\rho+\delta)\, e^{i\phi_j}, \quad
\phi_j=\frac{\pi j}{6}\,,\quad j=0,1,2,3\,, \label{num4}
\eeq
with $0<\rho<1$,\ $\delta=0.002$,
and calculated the corresponding values of $\Phiimh(z)$
from Eqs. \eqref{num1}  and \eqref{lab15}. For further reference, it is
convenient to split these points into the sets
\beq
\Delta_i=\{\>0.1(i-1)+\delta<|z-Z_0|<0.1\,i+\delta\,\}\label{num6d}\,,
\eeq
where $i=1,\ldots,10$, containing 200 points each.
All calculations were done with several values of $\tau$, given by Eq. 
\eqref{tauval}, for each value of $\xi$ (or $z$), as previously
explained. A difference between the values of the scaling function
with the same $\xi$ (but different $\tau$)
never exceeded $2\times10^{-10}$, so
we kept 10 decimal figures in our results for $\Phiimh(z)$.
Some numerical data for the function $\Phiimh(z)$
are given in the Appendix.

The first test for our numerical analysis is to demonstrate the presence of
the leading singular contribution
$(z-Z_0)^{5/6}$
in Eq. \eqref{num2} to confirm the CFT prediction \eqref{num2}. In doing
this, we need to suppress
the regular terms
$ a_{0,0}+a_{0,1}(z-Z_0)
$
which smoothen the behavior of $\Phiimh(z)$ near $z=Z_0$.
This is achieved by differentiating \eqref{num2}
with respect to $z$  and considering
$\mbox{Im}\Phiimh'(z)$ with $z=Z_0+i\epsilon,\ \epsilon>0$ (here the prime
denotes the $z$-derivative). Since we expect $a_{1,0}$ to be real, a
simple calculation gives
\beq \mbox{Im}\,\Phiimh'(Z_0+i \epsilon)=-\frac{5Z_0^2}{6}\,a_{1,0}\,
\sin\frac{\pi}{12}\,\epsilon^{-1/6}+
O(\epsilon^{2/3}). \label{num6}
\eeq

\begin{figure}
\begin{center}
\vspace{0.2cm}
    \includegraphics[width=0.9\columnwidth]{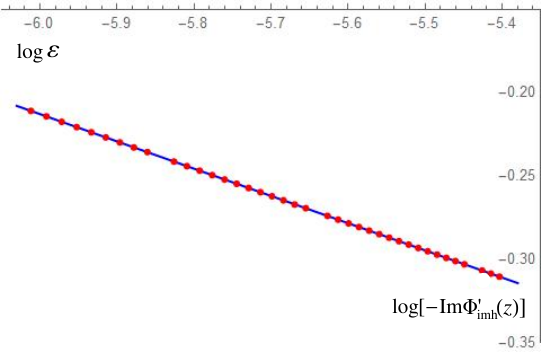}
\caption{Numerical data points (red dots) for
  $\log[-\mbox{Im}\Phiimh'(Z_0+i\epsilon)]$ fitted by Eq. \eqref{num6a}.
}
  \label{logphi}
  \end{center}
\end{figure}

Using a very accurate numerical differentiation, we have
calculated the data points for $\mbox{Im}\Phiimh'(z)$ in the range
$2.5\times 10^{-3}<\epsilon<4.5\times 10^{-3}$ and obtained the following fit
\begin{align}
&\log[-\mbox{Im}\,\Phiimh'(Z_0+i\epsilon)]=\nonumber\\
&-0.1650 \log(\epsilon)-1.204+0.2197\, \epsilon^{5/6}+O(\epsilon),\label{num6a}
\end{align}
which also takes into account the correction term in Eq. \eqref{num6}.
The fitting function \eqref{num6a} together with numerical data points
is shown in Fig.\ref{logphi} in the logarithmic scale.

As expected, the plot is very close to the straight line with the slope
$-\frac{1}{6}$ with a small deviation due to the next order
corrections in Eq. \eqref{num6}.
The result
confirms the CFT prediction for the exponent ($5/6$) of the leading
singular term in Eq. 
\eqref{num2} to within $0.2\%$ accuracy (our value is $5.010/6$).
For the subsequent fits, we used the nonlinear formula obtained from
Eqs. \eqref{gterm} and \eqref{Fsing}, with truncated series \eqref{alphaz} and
\eqref{gz} dropping all terms whose contributions into
Eq. \eqref{num2} are of the orders higher than $O\big[(z-Z_0)^8\big]$.
As follows from our estimates, the coefficients $a_{k,l}$ for such
terms appear be of the order $10^{-2}$ and smaller.
This means that for $|z-Z_0|<0.2$ contributions of the discarded terms
are expected to be of the order of $10^{-9}$ or less.

For this reason the main nonlinear fit for
$\Phiimh(z)$ we used
$400$ points \eqref{num4} with
$0<\rho<0.2$, i.e., for which $z\in(\Delta_1\cup\Delta_2)$, as defined
in Eq. \eqref{num6d}.
Subsequently, the accuracy of the fit
was tested for all subsets \eqref{num6d} of calculated data points.
The Table \ref{err} contains
the maximum difference (error) between the value of the fit
and a numerical value of $\Phiimh(z)$  calculated with the CTM method,
\beq
\Sigma_i= \max\limits_{z\in\Delta_i}|\Phiimh(z)-\Phiimh^{\mbox{\tiny CTM}}(z)|\,.
\eeq

\begin{widetext}
\onecolumngrid
\begin{table}[ht]
\caption{ The maximum error between $\Phiimh^{\mbox{\scriptsize trun}}(z)$ and
$\Phiimh^{\mbox{\tiny CTM}}(z)$
for the points $z\in\Delta_i$.}
\def\arraystretch{1.6}
\begin{tabular}{c|c|c|c|c|c|c|c|c|c|c}
\toprule   \vspace{-15pt}\\
$i$ & 1 & 2 & 3 & 4 & 5 & 6 & 7 & 8 & 9 & 10 \\
\hline
$\Sigma_{i}$& $ 1.1\times 10^{-10}$ & $4.3\times 10^{-11}$ & $1.8\times 10^{-8}$ & $3.0\times
   10^{-7}$ & $1.8\times 10^{-6}$ & $1.6\times 10^{-5}$ & $1.2\times 10^{-4}$ & $6.5\times 10^{-4}$ &
   $3.1\times 10^{-3}$ & $1.3\times 10^{-2}$ \\
   \botrule
\end{tabular}
\label{err}
\end{table}
\end{widetext}
\twocolumngrid
Since the points $z\in\{\Delta_1\cup\Delta_2\}$ were used for the fit,
it is not surprising that for $i=1,2$ the error is of the order of $10^{-10}$ or less.
In fact, it coincides
with the accuracy of our numerical calculations of $\Phiimh(z)$.
The same accuracy also holds for the location of the singularity
\beq
Z_0=2.4291691718(2). \label{num7}
\eeq
arising from our final fit.
The corresponding position of the singular point $\xi_0$ for $\ghigh(\xi)$ is
\beq
\xi_0=0.18935060551(3)\,.\label{num8}
\eeq
Further, Table
\ref{err} shows that for the subsequent sets $z\in\Delta_i$, with $i\ge 3$,
the accuracy of the fit drops down, which is natural
due to an increasing effect
of truncation of the series \eqref{num2}. Nevertheless, the value of $Z_0$,
given in Eq. \eqref{num7}, remains stable, even
if points $z\in\Delta_i$, with $i\ge 3$,
are included in the fit.

\begin{table}[ht]
\caption{Coefficients $\hat g_k$, $\hat\lambda_k$ and $\hat\alpha_k$ }
\def\arraystretch{1.4}
\begin{center}
\begin{tabular}{|c|c|c|c|}
\hline
\phantom{a}k\phantom{a} & $\hat g_k$ & $\hat\lambda_k$& $\hat\alpha_k$ \\
\hline
0 &0.0928378351 & ~---&-1.3478038278\\
\hline
1 &0.1035665061& 0.1720881869& 0.0845043112\\
\hline
2&-0.0406336255&-0.0530975348&-0.1988535226\\
\hline
3&-0.1021530341&-0.0432647893& 0.1096925182\\
\hline
4& 0.0622170621&-0.0060926724& 0.0588155481\\
\hline
5&-0.0427391325& 0.0015477546 &-0.0907062356\\
\hline
6& 0.0367166751&-0.0026569648&0.0650203209\\
\hline
7&0.0120039152&-0.0046983681&~---\\
\hline
8&-0.0179189367 &-0.0050775498&~---\\
\hline
\end{tabular}
\end{center}
\label{lamdata}
\end{table}

The final fit for the coefficients $\hat g_k$, $\hat\lambda_k$ and $\hat\alpha_k$ from
Eqs. \eqref{alphaz} and \eqref{gz} is
shown in Table \ref{lamdata}.  Let us notice that the higher
coefficients may be correct only up to one or two significant
digits due to the series truncation.  The lower coefficients should be still
correct up to eight or nine digits. It is hard to estimate the number of
correct digits in each coefficient, so we left them as they were produced
by {\it Mathematica}.

Let us comment on this in more detail.
We need to use all digits in  coefficients in Table \ref{lamdata} to produce accurate
values for $\Phiimh(z)$ from the Appendix.
Any truncation will cause a significant deviation from the values given there.
It seems that this contradicts the statement from the previous paragraph.
The explanation is that the series \eqref{num2} contains too many independent powers in
$v$ with a very small separation of exponents.
As a result, one can construct another fit for $\Phiimh(z)$
which will reproduce the data from Appendix with $10^{-10}$ accuracy but
higher coefficients will coincide only up to one or two significant digits. In other words,
one needs the
data with much higher precision to get reliable information about
coefficients with $k>2$ in Table \ref{lamdata}.

Now let us give coefficients $a_{k,l}$ of the series expansions \eqref{num2} for the function
$\gly(z)$. They are shown in Table \ref{phidata}. The coefficient $b_0$ coming from the
leading power in the second term of Eq. \eqref{exp1} is estimated as
\beq
b_0=0.0623126407.
\eeq
Let us notice that to get the series expansion for $\Phiimh(z)$ from $\gly(z)$ we could use Eq. \eqref{num1}.
However, due to the presence of the extra factor $(Z_0+v)^2$, coefficients
for $\Phiimh(z)$ will be in the range $(0.1,0.5)$.

We can also get an estimate for the term $c_{\Xi}\,\beta(z)$ in \eqref{exp1}
\begin{align}
&c_{\Xi}\,\beta(z)=0.01015847 + 0.0070108\, v\nonumber\\
&+ 0.009344\, v^2 +
 0.006279\, v^3 - 0.00058\, v^4 +\ldots \label{cbeta1}
\end{align}
where $v=z-Z_0$.
From the third term in Eq. \eqref{exp1} we can estimate $c_{\Xi\Xi}$
\beq
c_{\Xi\Xi}\approx 7.69\, c_{\Xi}^2.\label{cbeta2}
\eeq
Finally, from the expansion for the fourth term in Eq. \eqref{num2} we obtain
\beq
c_{\Xi_8}\delta(z)=-0.00064 - 0.00094\, (z-Z_0) +\ldots \label{cbeta3}
\eeq
It appears, that the values in Eqs. (\ref{cbeta1}-\ref{cbeta3}) should
be understood only as an order-of-magnitude estimate.

\begin{widetext}
\onecolumngrid
\begin{table}[ht]
\caption{Coefficients $a_{k,l}$ of the truncated scaling function $\gly(z)$ from \eqref{num2}}
\def\arraystretch{1.4}
\begin{center}
\begin{tabular}{|*{7}{c|}}
\hline
 \diagbox{l}{k}& 0 & 1& 2&3& 4& 5\\
 \cline{1-7}
0&0.0928378351& -0.2326384010& 0.0729439864& -0.0228716546&  0.0071714285& -0.0201278728\\ \cline{1-7}
1&0.1035665061& 0.0598168387& -0.0420846767&  0.0205105351& -0.0087246739& -0.0105062255\\ \cline{1-7}
2&-0.0406336255&  0.0502778268& -0.0135928541& 0.0012416359& 0.0012912732& -0.0151031033\\ \cline{1-7}
3&-0.1021530341&  0.0095546595& -0.0076813505&  0.0222269546& -0.0081326053& -0.036304106\\ \cline{1-7}
4&0.0231406896&  0.0184911041& -0.0070095153& -0.0314395832&  0.0212048796\\ \cline{1-6}
5&-0.0183244972& -0.0135214010& 0.0173465826&  0.0001177947\\ \cline{1-5}
6&0.0426269803&  0.0054321576& -0.0088442156\\ \cline{1-4}
7&0.0237701200& -0.0002697285\\ \cline{1-3}
8&0.0032474000\\ \cline{1-2}
\end{tabular}
\end{center}
\label{phidata}
\end{table}
\end{widetext}
\twocolumngrid

Note that the value of $a_{1,0}$ can also be estimated from Eq. \eqref{num6a}
\beq
a_{1,0}\approx -\frac{6}{5Z_0^2\sin(\pi/12)}\exp(-1.204)\approx -0.236, \label{num6b}
\eeq
which is below the more accurate value
\beq
a_{1,0}=2\fyl\cyl^2\hat\lambda_1^{5/6}=-0.232638\ldots\label{a1000}
\eeq
 The discrepancy is mostly caused
by the fact that Eq. \eqref{num6a} contains an approximate ``fitted''
value of the leading coefficient $-1/6$, while in Eq. \eqref{num2}
the related exponent $5/6$ is set to its exact value.

Next, we present a few coefficient functions $G_k(u)$ in the expansion
of $\ghigh(\xi)$ near $\xi=\pm i \xi_0$ in terms of variable $u=\xi^2+\xi_0^2$,
\beq
\gly(\xi)=G_0(u)+G_1(u)u^{\frac{5}{6}}+
G_2(u)u^{\frac{5}{3}}+G_3(u)u^{\frac{5}{2}}+
\cdots
\eeq
\beq
\begin{array}{ll}
G_0(u)&=0.09284+1.871u+19.8u^2-371u^3+\ldots\\[.1cm]
G_1(u)&=-2.5947-26.154u-172u^2+\ldots\\[.1cm]
G_2(u)&=9.076+171.35u+2091u^2+\ldots,\\[.1cm]
G_3(u)&=-31.75-878.7u+\ldots.\label{num13}
\end{array}
\eeq

\begin{figure}[ht]
\begin{center}
    \includegraphics[width=0.9\columnwidth]{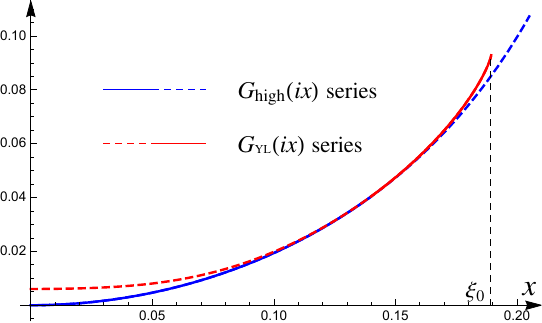}
\caption{A comparison of $\ghigh(\xi)$ and $\gly(\xi)$ along the imaginary axis $\xi=ix$}
  \label{compare}
  \end{center}
\end{figure}

It is instructive to compare the previous results \cite{MBBD09}
for the scaling function
$\ghigh(\xi)$ [given in Eq. \eqref{lab8} and Table \ref{tablef}]
and our new results for the function $\gly(\xi)$
for purely imaginary values of the
argument $\xi=ix$ in the interval  $x\in(0,0.2)$.
As shown in Fig. \ref{compare}, the two functions match
very well in the intermediate region and deviate from each other
towards the ends of the interval.

Finally, note that our results for the leading coefficients of the
coupling constant
\beq
\lambda(z)=\hat\lambda_1 \, v+\hat\lambda_2\,v^2+\ldots =
\lambda_1 \, u+\lambda_2\,u^2+\ldots
\eeq
are in excellent agreement with the results of Refs. \cite{FZ03,XuZ22}
[the variables $v$ and $u$ are defined in Eqs. \eqref{x-def} and \eqref{u-def}]. Indeed,
our values
\beq
\lambda_1=3.10916 (2)\,,\qquad \lambda_2=37.6 (8)\,,
\eeq
corresponding to $\hat\lambda_1,\hat\lambda_2$ from Table~\ref{lamdata}, perfectly match the values
$\lambda_1=3.089\pm0.008\,,\ \lambda_2=38.4\pm1.6$, given by Eq. (4.7) of
Ref. \cite{XuZ22}.
Similarly, their values for $f_0=0.092746\ldots$ (denoted here as $\hat g_0$)
and for $\hat\alpha_0=-1.32\pm0.05$, given in their eqs. (3.10) and (4.8), respectively,
 match our values presented in Table~\ref{lamdata}.

\section{Conclusion\label{secconc}}
One of the motivations of our work was to confirm and extend the field
theory results \cite{FZ03,XuZ22} on the Yang-Lee edge singularity
through {\it ab initio} calculations, directly from
the original lattice formulation \eqref{Z-def} of the Ising model.

We used Baxter's variational
approach based on the corner transfer matrix (CTM) method \cite{Baxter2007},
enhanced by an
improved iteration scheme \cite{Nishino97}, known as the corner transfer
matrix renormalization group (CTMRG).
The main advantage of this approach over other numerical schemes (e.g., the
row-to-row transfer matrix method) is that it is formulated
directly in the limit of an {\em infinite} lattice.
Its accuracy depends on the magnitude of truncated eigenvalues of the corner
transfer matrix (which is at our control), rather than the size of the
lattice. This allows one to calculate the lattice free energy of the model
with a rather high precision, of about 30 digits.
However, the accuracy of the calculation of the
universal characteristics of the continuous scaling theory
is limited by the existence of unknown
lattice-dependent subleading contribution to the free energy.
This is the reason we have chosen the triangular lattice,
where such contributions only appear in the order of $\tau^6$, where
$\tau\sim (T-T_c)$ is the deviation from the lattice critical temperature. Next,
we were able to completely determine these $O(\tau^6)$ contributions
by using the best available perturbation theory calculation of the
magnetic susceptibility of the lattice Ising model \cite{CGNP11}.

In the analysis presented in Sec.\ref{sectria} and
Sec.\ref{secsus} we used all available exact and perturbation
theory results on the 2D lattice Ising model to determine the
lattice-dependent regular and subleading contributions to the free
energy, as well as to find coefficients
entering the nonlinear scaling variables \eqref{lab4} to
highest possible orders in the variables $\tau$ and $H$.
After these preparations the universal part of
the free energy has been numerically calculated using Eq.
\eqref{lab15}
for a large number of different values of the scaling parameter $\xi$
(or $z$),   defined in Eq. \eqref{zvardef}.
The use of the nonlinear
scaling variables \eqref{lab4} allowed us to perform calculations
sufficiently away from the critical point with a reliable convergence
of the algorithm. Using this technique
we have numerically calculated (with 10-digit accuracy) the universal
scaling function of the Ising Field Theory in the vicinity of the
Yang-Lee singularity. This data was used to numerically find a set of main
parameters describing the Ising Field Theory
as a model of perturbed minimal conformal field theory $M_{2/5}$
involving an infinite tower of irrelevant operators.
We also determined the location of the Yang-Lee singularity with
a much higher (10 digits) accuracy than it was previously known
\cite{XuZ22,XuZ23} and
confirmed the leading exponent in the singular expansion of the free
energy near the critical point predicted by the conformal field theory.
\section{Acknowledgments
}
The authors thank Prof. F. Smirnov for important comments
and Prof. A. Zamolodchikov for extremely useful
discussions at various stages of our study. V.V.M. acknowledges
the support of the Australian Research Council, Grant
No. DP190103144.

\bibliographystyle{apsrev4-2}

\newcommand\oneletter[1]{#1}

\pagebreak
\widetext
\begin{center}
\textbf{\large Appendix: Numerical data for the scaling function}
\end{center}
\setcounter{equation}{0}
\setcounter{figure}{0}
\setcounter{page}{1}
\makeatletter
\renewcommand{\theequation}{S\arabic{equation}}
\renewcommand{\thefigure}{S\arabic{figure}}
\renewcommand{\bibnumfmt}[1]{[S#1]}
\renewcommand{\citenumfont}[1]{S#1}
In this Appendix we give numerical values for the function
$\Phiimh^{\mbox{\tiny CTM}}(z)$
in the regions $\Delta_1$, $\Delta_2$ and $\Delta_3$ from \eqref{num6d}.
In our original fitting we used the step $0.002$ for values of $\rho$.
Here we only give 50\% of all points
and use the step $0.004$. The values of $z$ are given by
\beq
z_{kj}=Z_0^{\mbox{\scriptsize in}}+(0.0001+0.004k)e^{\frac{i\pi j}{6}},
\quad  k=1,\ldots,75,\quad j=0,\ldots,3 \label{S1}
\eeq
\beq
Z_0^{\mbox{\scriptsize in}}=2.4291691718. \label{S2}
\eeq
In Table \ref{tabdata1} all points belong to the region $\Delta_1\cup\Delta_2$.
The maximum error for these data is $1.87\times 10^{-10}$.
The error is estimated through the calculation of $\Phiimh^{\mbox{\tiny CTM}}(z)$ for several
values of $\tau$ as explained after \eqref{num6d}.
It is slightly higher than the error $\Sigma_1$ in the first
column of Table \ref{err}.
In Table \ref{tabdata2} we give values for points in $\Delta_3$.

\bgroup
\begin{table}[ht]
\caption{ The values of the function $\Phiimh^{\mbox{\tiny CTM}}(z_{kj})$
for $k=1\ldots,50$ in \eqref{S1}}
	\begin{center}
\def\arraystretch{1.2}
\begin{tabular}{|*{5}{c|}}
\hline
k & j=0 & j=1 & j=2 & j=3 \\ \hline
 1 & 0.5381281124 & 0.5388576178-0.0037551981$\,i$ & 0.5409580608-0.0069867366$\,i$ &
   0.5441685923-0.0092201138$\,i$ \\ \hline
 2 & 0.5316643828 & 0.5328233012-0.0061499721$\,i$ & 0.5361791624-0.0115048992$\,i$ &
   0.5413665298-0.0153222565$\,i$ \\ \hline
 3 & 0.5260966328 & 0.5276005678-0.0081648500$\,i$ & 0.5319740744-0.0153345159$\,i$ &
   0.5387912192-0.0205573761$\,i$ \\ \hline
 4 & 0.5210726235 & 0.5228705683-0.0099492992$\,i$ & 0.5281177055-0.0187460166$\,i$ &
   0.5363536754-0.0252653491$\,i$ \\ \hline
 5 & 0.5164343051 & 0.5184902651-0.0115704462$\,i$ & 0.5245091767-0.0218608258$\,i$ &
   0.5340142086-0.0295986922$\,i$ \\ \hline
 6 & 0.5120922803 & 0.5143787623-0.0130662842$\,i$ & 0.5210914732-0.0247476642$\,i$ &
   0.5317504712-0.0336436922$\,i$ \\ \hline
 7 & 0.5079893267 & 0.5104842833-0.0144611440$\,i$ & 0.5178281207-0.0274505408$\,i$ &
   0.5295482110-0.0374555959$\,i$ \\ \hline
 8 & 0.5040858306 & 0.5067709856-0.0157718755$\,i$ & 0.5146938778-0.0299999370$\,i$ &
   0.5273975912-0.0410726638$\,i$ \\ \hline
 9 & 0.5003528306 & 0.5032126646-0.0170108013$\,i$ & 0.5116702953-0.0324181461$\,i$ &
   0.5252914378-0.0445228788$\,i$ \\ \hline
 10 & 0.4967682841 & 0.4997893762-0.0181873018$\,i$ & 0.5087433332-0.0347221428$\,i$ &
   0.5232243020-0.0478275482$\,i$ \\ \hline
 11 & 0.4933148872 & 0.4964854624-0.0193087421$\,i$ & 0.5059019694-0.0369252573$\,i$ &
   0.5211919134-0.0510034067$\,i$ \\ \hline
 12 & 0.4899787168 & 0.4932883244-0.0203810488$\,i$ & 0.5031373353-0.0390382187$\,i$ &
   0.5191908407-0.0540639260$\,i$ \\ \hline
 13 & 0.4867483441 & 0.4901876190-0.0214090870$\,i$ & 0.5004421500-0.0410698372$\,i$ &
   0.5172182704-0.0570201703$\,i$ \\ \hline
 14 & 0.4836142303 & 0.4871747133-0.0223969170$\,i$ & 0.4978103353-0.0430274680$\,i$ &
   0.5152718561-0.0598813788$\,i$ \\ \hline
 15 & 0.4805683018 & 0.4842422993-0.0233479759$\,i$ & 0.4952367452-0.0449173385$\,i$ &
   0.5133496129-0.0626553764$\,i$ \\ \hline
 16 & 0.4776036425 & 0.4813841164-0.0242652072$\,i$ & 0.4927169705-0.0467447846$\,i$ &
   0.5114498414-0.0653488693$\,i$ \\ \hline
 17 & 0.4747142661 & 0.4785947448-0.0251511589$\,i$ & 0.4902471931-0.0485144256$\,i$ &
   0.5095710707-0.0679676661$\,i$ \\ \hline
 18 & 0.4718949444 & 0.4758694503-0.0260080558$\,i$ & 0.4878240772-0.0502302976$\,i$ &
   0.5077120165-0.0705168428$\,i$ \\ \hline
 19 & 0.4691410739 & 0.4732040640-0.0268378567$\,i$ & 0.4854446843-0.0518959546$\,i$ &
   0.5058715475-0.0730008712$\,i$ \\ \hline
 20 & 0.4664485728 & 0.4705948890-0.0276422980$\,i$ & 0.4831064078-0.0535145486$\,i$ &
   0.5040486604-0.0754237191$\,i$ \\ \hline
 21 & 0.4638137990 & 0.4680386259-0.0284229288$\,i$ & 0.4808069204-0.0550888928$\,i$ &
   0.5022424597-0.0777889290$\,i$ \\ \hline
 22 & 0.4612334838 & 0.4655323133-0.0291811392$\,i$ & 0.4785441326-0.0566215125$\,i$ &
   0.5004521412-0.0800996825$\,i$ \\ \hline
 23 & 0.4587046785 & 0.4630732792-0.0299181827$\,i$ & 0.4763161585-0.0581146859$\,i$ &
   0.4986769789-0.0823588509$\,i$ \\ \hline
 24 & 0.4562247111 & 0.4606591018-0.0306351951$\,i$ & 0.4741212875-0.0595704781$\,i$ &
   0.4969163144-0.0845690386$\,i$ \\ \hline
 25 & 0.4537911490 & 0.4582875763-0.0313332100$\,i$ & 0.4719579624-0.0609907693$\,i$ &
   0.4951695480-0.0867326173$\,i$ \\ \hline
 26 & 0.4514017700 & 0.4559566876-0.0320131714$\,i$ & 0.4698247587-0.0623772773$\,i$ &
   0.4934361308-0.0888517558$\,i$ \\ \hline
 27 & 0.4490545358 & 0.4536645878-0.0326759450$\,i$ & 0.4677203696-0.0637315780$\,i$ &
   0.4917155594-0.0909284440$\,i$ \\ \hline
 28 & 0.4467475713 & 0.4514095759-0.0333223268$\,i$ & 0.4656435919-0.0650551216$\,i$ &
   0.4900073696-0.0929645144$\,i$ \\ \hline
 29 & 0.4444791458 & 0.4491900821-0.0339530514$\,i$ & 0.4635933143-0.0663492465$\,i$ &
   0.4883111328-0.0949616591$\,i$ \\ \hline
 30 & 0.4422476574 & 0.4470046528-0.0345687985$\,i$ & 0.4615685075-0.0676151917$\,i$ &
   0.4866264515-0.0969214452$\,i$ \\ \hline
 31 & 0.4400516192 & 0.4448519390-0.0351701985$\,i$ & 0.4595682156-0.0688541072$\,i$ &
   0.4849529565-0.0988453284$\,i$ \\ \hline
 32 & 0.4378896486 & 0.4427306853-0.0357578378$\,i$ & 0.4575915486-0.0700670629$\,i$ &
   0.4832903036-0.1007346635$\,i$ \\ \hline
 33 & 0.4357604556 & 0.4406397209-0.0363322630$\,i$ & 0.4556376760-0.0712550567$\,i$ &
   0.4816381714-0.1025907149$\,i$ \\ \hline
 34 & 0.4336628353 & 0.4385779517-0.0368939847$\,i$ & 0.4537058210-0.0724190213$\,i$ &
   0.4799962592-0.1044146650$\,i$ \\ \hline
 35 & 0.4315956590 & 0.4365443522-0.0374434807$\,i$ & 0.4517952559-0.0735598299$\,i$ &
   0.4783642848-0.1062076215$\,i$ \\ \hline
 36 & 0.4295578677 & 0.4345379609-0.0379811995$\,i$ & 0.4499052971-0.0746783020$\,i$ &
   0.4767419829-0.1079706247$\,i$ \\ \hline
 37 & 0.4275484663 & 0.4325578734-0.0385075620$\,i$ & 0.4480353017-0.0757752080$\,i$ &
   0.4751291039-0.1097046528$\,i$ \\ \hline
 38 & 0.4255665178 & 0.4306032380-0.0390229647$\,i$ & 0.4461846638-0.0768512731$\,i$ &
   0.4735254122-0.1114106278$\,i$ \\ \hline
 39 & 0.4236111385 & 0.4286732516-0.0395277812$\,i$ & 0.4443528118-0.0779071816$\,i$ &
   0.4719306851-0.1130894193$\,i$ \\ \hline
 40 & 0.4216814936 & 0.4267671553-0.0400223642$\,i$ & 0.4425392049-0.0789435794$\,i$ &
   0.4703447122-0.1147418496$\,i$ \\ \hline
 41 & 0.4197767938 & 0.4248842310-0.0405070472$\,i$ & 0.4407433317-0.0799610779$\,i$ &
   0.4687672937-0.1163686968$\,i$ \\ \hline
 42 & 0.4178962913 & 0.4230237988-0.0409821459$\,i$ & 0.4389647071-0.0809602558$\,i$ &
   0.4671982402-0.1179706984$\,i$ \\ \hline
 43 & 0.4160392770 & 0.4211852132-0.0414479597$\,i$ & 0.4372028704-0.0819416625$\,i$ &
   0.4656373716-0.1195485544$\,i$ \\ \hline
 44 & 0.4142050774 & 0.4193678617-0.0419047727$\,i$ & 0.4354573842-0.0829058193$\,i$ &
   0.4640845166-0.1211029299$\,i$ \\ \hline
 45 & 0.4123930522 & 0.4175711612-0.0423528547$\,i$ & 0.4337278321-0.0838532219$\,i$ &
   0.4625395118-0.1226344578$\,i$ \\ \hline
 46 & 0.4106025921 & 0.4157945570-0.0427924627$\,i$ & 0.4320138174-0.0847843423$\,i$ &
   0.4610022015-0.1241437408$\,i$ \\ \hline
 47 & 0.4088331163 & 0.4140375204-0.0432238413$\,i$ & 0.4303149616-0.0856996303$\,i$ &
   0.4594724370-0.1256313534$\,i$ \\ \hline
 48 & 0.4070840710 & 0.4122995468-0.0436472237$\,i$ & 0.4286309036-0.0865995147$\,i$ &
   0.4579500760-0.1270978444$\,i$ \\ \hline
 49 & 0.4053549275 & 0.4105801545-0.0440628325$\,i$ & 0.4269612983-0.0874844051$\,i$ &
   0.4564349824-0.1285437377$\,i$ \\ \hline
 50 & 0.4036451802 & 0.4088788829-0.0444708802$\,i$ & 0.4253058155-0.0883546928$\,i$ &
   0.4549270260-0.1299695344$\,i$ \\ \hline
\end{tabular}
\end{center}
\label{tabdata1}
\end{table}
\egroup

\bgroup
\begin{table}[ht]
\caption{ The values of the function $\Phiimh^{\mbox{\tiny CTM}}(z_{kj})$
for $k=51,\ldots,75$ $\,i$n \eqref{S1}}
	\begin{center}
\def\arraystretch{1.2}
\begin{tabular}{|*{5}{c|}}
\hline
k & j=0 & j=1 & j=2 & j=3 \\ \hline
 51 & 0.4019543458 & 0.4071952913-0.0448715701$\,i$ & 0.4236641390-0.0892107524$\,i$ &
   0.4534260817-0.1313757145$\,i$ \\ \hline
 52 & 0.4002819614 & 0.4055289577-0.0452650967$\,i$ & 0.4220359659-0.0900529422$\,i$ &
   0.4519320294-0.1327627374$\,i$ \\ \hline
 53 & 0.3986275835 & 0.4038794777-0.0456516463$\,i$ & 0.4204210056-0.0908816059$\,i$ &
   0.4504447541-0.1341310440$\,i$ \\ \hline
 54 & 0.3969907869 & 0.4022464634-0.0460313975$\,i$ & 0.4188189794-0.0916970728$\,i$ &
   0.4489641450-0.1354810570$\,i$ \\ \hline
 55 & 0.3953711632 & 0.4006295423-0.0464045214$\,i$ & 0.4172296190-0.0924996591$\,i$ &
   0.4474900954-0.1368131827$\,i$ \\ \hline
 56 & 0.3937683205 & 0.3990283567-0.0467711826$\,i$ & 0.4156526673-0.0932896684$\,i$ &
   0.4460225027-0.1381278114$\,i$ \\ \hline
 57 & 0.3921818820 & 0.3974425624-0.0471315389$\,i$ & 0.4140878761-0.0940673926$\,i$ &
   0.4445612681-0.1394253186$\,i$ \\ \hline
 58 & 0.3906114851 & 0.3958718287-0.0474857422$\,i$ & 0.4125350071-0.0948331124$\,i$ &
   0.4431062961-0.1407060656$\,i$ \\ \hline
 59 & 0.3890567810 & 0.3943158370-0.0478339386$\,i$ & 0.4109938303-0.0955870981$\,i$ &
   0.4416574949-0.1419704005$\,i$ \\ \hline
 60 & 0.3875174339 & 0.3927742803-0.0481762685$\,i$ & 0.4094641243-0.0963296098$\,i$ &
   0.4402147755-0.1432186585$\,i$ \\ \hline
 61 & 0.3859931199 & 0.3912468629-0.0485128675$\,i$ & 0.4079456753-0.0970608982$\,i$ &
   0.4387780522-0.1444511633$\,i$ \\ \hline
 62 & 0.3844835269 & 0.3897332993-0.0488438661$\,i$ & 0.4064382769-0.0977812052$\,i$ &
   0.4373472419-0.1456682267$\,i$ \\ \hline
 63 & 0.3829883536 & 0.3882333142-0.0491693900$\,i$ & 0.4049417301-0.0984907641$\,i$ &
   0.4359222643-0.1468701502$\,i$ \\ \hline
 64 & 0.3815073092 & 0.3867466418-0.0494895607$\,i$ & 0.4034558422-0.0991898001$\,i$ &
   0.4345030418-0.1480572248$\,i$ \\ \hline
 65 & 0.3800401129 & 0.3852730251-0.0498044954$\,i$ & 0.4019804272-0.0998785308$\,i$ &
   0.4330894990-0.1492297318$\,i$ \\ \hline
 66 & 0.3785864931 & 0.3838122157-0.0501143074$\,i$ & 0.4005153050-0.1005571663$\,i$ &
   0.4316815631-0.1503879434$\,i$ \\ \hline
 67 & 0.3771461874 & 0.3823639734-0.0504191060$\,i$ & 0.3990603012-0.1012259101$\,i$ &
   0.4302791634-0.1515321227$\,i$ \\ \hline
 68 & 0.3757189420 & 0.3809280659-0.0507189970$\,i$ & 0.3976152471-0.1018849586$\,i$ &
   0.4288822313-0.1526625247$\,i$ \\ \hline
 69 & 0.3743045108 & 0.3795042680-0.0510140828$\,i$ & 0.3961799791-0.1025345023$\,i$ &
   0.4274907001-0.1537793961$\,i$ \\ \hline
 70 & 0.3729026561 & 0.3780923618-0.0513044624$\,i$ & 0.3947543389-0.1031747255$\,i$ &
   0.4261045053-0.1548829760$\,i$ \\ \hline
 71 & 0.3715131472 & 0.3766921361-0.0515902317$\,i$ & 0.3933381726-0.1038058068$\,i$ &
   0.4247235841-0.1559734963$\,i$ \\ \hline
 72 & 0.3701357607 & 0.3753033861-0.0518714835$\,i$ & 0.3919313311-0.1044279190$\,i$ &
   0.4233478755-0.1570511819$\,i$ \\ \hline
 73 & 0.3687702797 & 0.3739259134-0.0521483079$\,i$ & 0.3905336697-0.1050412301$\,i$ &
   0.4219773200-0.1581162509$\,i$ \\ \hline
 74 & 0.3674164943 & 0.3725595249-0.0524207921$\,i$ & 0.3891450479-0.1056459029$\,i$ &
   0.4206118602-0.1591689151$\,i$ \\ \hline
 75 & 0.3660742004 & 0.3712040338-0.0526890207$\,i$ & 0.3877653291-0.1062420951$\,i$ &
   0.4192514396-0.1602093800$\,i$ \\ \hline
\end{tabular}
\end{center}
\label{tabdata2}
\end{table}
\egroup

\end{document}